%% file: NoParticleProduction_v6.tex
\documentclass[12pt]{article}

\pdfoutput=1
\usepackage[colorlinks=true,linkcolor=darkblue,citecolor=darkblue,urlcolor=darkblue]{hyperref}
\usepackage[T1]{fontenc}
\usepackage[ansinew]{inputenc}
\usepackage[english]{babel}
\usepackage{color}
\usepackage{epsfig, palatino}
\usepackage{ae}
\usepackage{ulem} 
\usepackage{amsmath}
\usepackage{amssymb}
\usepackage{tikz}
\usepackage{graphicx}
\usepackage[export]{adjustbox}
\usepackage{color}
\definecolor{darkblue}{cmyk}{0.9,0.9,0,0}
\usepackage{cite}
\usepackage{wasysym}
\usepackage{varioref}
\usepackage{makeidx}
\usepackage{simplewick}
\usepackage{dsfont}
\usepackage{slashed}

\DeclareMathOperator*{\res}{Res}

\makeindex

\numberwithin{figure}{section}

\begin{document}

\thispagestyle{empty}

\setcounter{page}{1}
\setcounter{footnote}{0}
\setcounter{figure}{0}
\begin{center}
$$$$
{\Large\textbf{\mathversion{bold}
No Particle Production in Two Dimensions:\\
Recursion Relations and Multi-Regge Limit
}\par}

\vspace{1.0cm}
Barak Gabai$^{1,2}$, Dalimil Maz\'{a}\v{c}$^{1,3,4}$, Andrei Shieber$^{1}$, Pedro Vieira$^{1,5}$, Yehao Zhou$^{1}$
\\ \vspace{1.2cm}
$^\text{\tiny 1}$Perimeter Institute for Theoretical Physics,
Waterloo, Ontario N2L 2Y5, Canada\\
$^\text{\tiny 2}$Jefferson Physical Laboratory, Harvard University, Cambridge, MA 02138, USA\\
$^\text{\tiny 3}$C. N. Yang Institute for Theoretical Physics, SUNY, Stony Brook, NY 11794, USA\\
$^\text{\tiny 4}$Simons Center for Geometry and Physics, SUNY, Stony Brook, NY 11794, USA\\
$^\text{\tiny 5}$ICTP South American Institute for Fundamental Research, IFT-UNESP, S\~ao Paulo, SP Brazil 01440-070  
\vspace{4mm}
\vspace{4mm}

\par\vspace{1.5cm}

\textbf{Abstract}\vspace{2mm}
\end{center}

We introduce high-energy limits which allow us to derive recursion relations fixing the various couplings of Lagrangians of two-dimensional relativistic quantum field theories with no tree-level particle production in a very straightforward way. The sine-Gordon model, the Bullough-Dodd theory, Toda theories of various kinds and the U(N) non-linear sigma model can all be rediscovered in this way. The results here were the outcome of our explorations at the 2017 Perimeter Institute Winter School.

\noindent

\newpage

\tableofcontents

\parskip 5pt plus 1pt   \jot = 1.5ex

\newpage

\section{Introduction, multi-Regge limit and recursion relations}

In the real world, collisions of particles at high enough energy produce additional particles.  For example, the amplitude~$\mathcal{M}_{2\to 4}$ for producing four particles out of two is non-zero in a generic relativistic quantum field theory. Indeed, this amplitude is related by crossing symmetry -- implemented by an analytic continuation -- to the amplitude $\mathcal{M}_{3\to 3}$ for three particles to evolve into three particles and the later is typically nonzero in a generic kinematical configuration. In two space-time dimensions, the so-called integrable theories constitute an important loophole to this statement. In these theories $\mathcal{M}_{3\to 3}$ is localized to a measure zero subspace of the kinematical space (corresponding to factorized scattering) while $\mathcal{M}_{2\to 4}$ vanishes identically. In higher dimensions, such measure-zero theories are necessarily free; i.e. if 
$\mathcal{M}_{3\to3}$ is non-trivial then so is $\mathcal{M}_{2\to4}$. At the same time -- without further physical input -- it is hard to rule out a very small but non-vanishing $\mathcal{M}_{2\to 4}$.

A recent motivation for studying theories with such very small particle production comes from the S-matrix bootstrap explorations of \cite{bootstrap1,bootstrap2}. In these works, the space of massive relativistic quantum field theories is carved out by looking for the maximal couplings between various physical particles given a fixed mass spectrum. This search was performed analytically in two dimensions \cite{bootstrap1} and it was found that the theories which maximize various couplings have no particle production. In higher dimensions, one has to resort to a numerical search \cite{bootstrap2} to find very little particle production also in this case. 

This motivated us to explore the space of quantum field theories without particle production. Concretely, we will consider massive scalars in two dimensions inspired by a beautiful review article \cite{Dorey:1996gd}, where Patrick Dorey points out that already at tree level, one can severely constrain the Lagrangians of two-dimensional quantum field theories by  imposing absence of particle production recursively. He illustrates how to carry out the first few steps of this recursive program to recover the first few terms in the expansion of the sine-Gordon and the Bullough-Dodd Lagrangians. These games are probably well known to the experts and date back all the way to a beautiful paper by I. Arefeva and V. Korepin in 1974 \cite{Arefeva:1974bk} where they first point out these tree level cancellations for the sine-Gordon model (and even considered the quantum version of these cancelations). In \cite{Braden:1991vz} more complicated theories were considered along the same lines. What has never been done -- as far as we are aware -- is a complete analysis of this recursive procedure which leads to the full form of these Lagrangians. This is what we set out to do during the 2017 Perimeter Institute Winter School.\footnote{We are grateful to all the PSI fellows and especially to Tibra Ali and Erica Goss for oranizing such a wonderful school.} In this short note, we present the outcome of this exercise. 

Let us present the gist of the argument. Consider for simplicity a single real scalar in (1+1)D with mass $m$ and interaction Lagrangian 
\begin{equation}
\mathcal{L}_\text{interaction}= -m^2 \sum_{n=3}^\infty \frac{v_n}{n!}\phi^n\,.
\end{equation}
The first production amplitude we want to suppress is $\mathcal{M}_{2\to 3}$. Setting all particles as incoming, and using the light-cone coordinates $p_{j} = m(a_j,1/a_j) $ we have
\begin{equation}
-\frac{1}{m^2}\mathcal{M}_{2\to 3} = \underbrace{v_3 v_4 \sum_{\alpha} G(\alpha) }_{\includegraphics[scale=0.06]{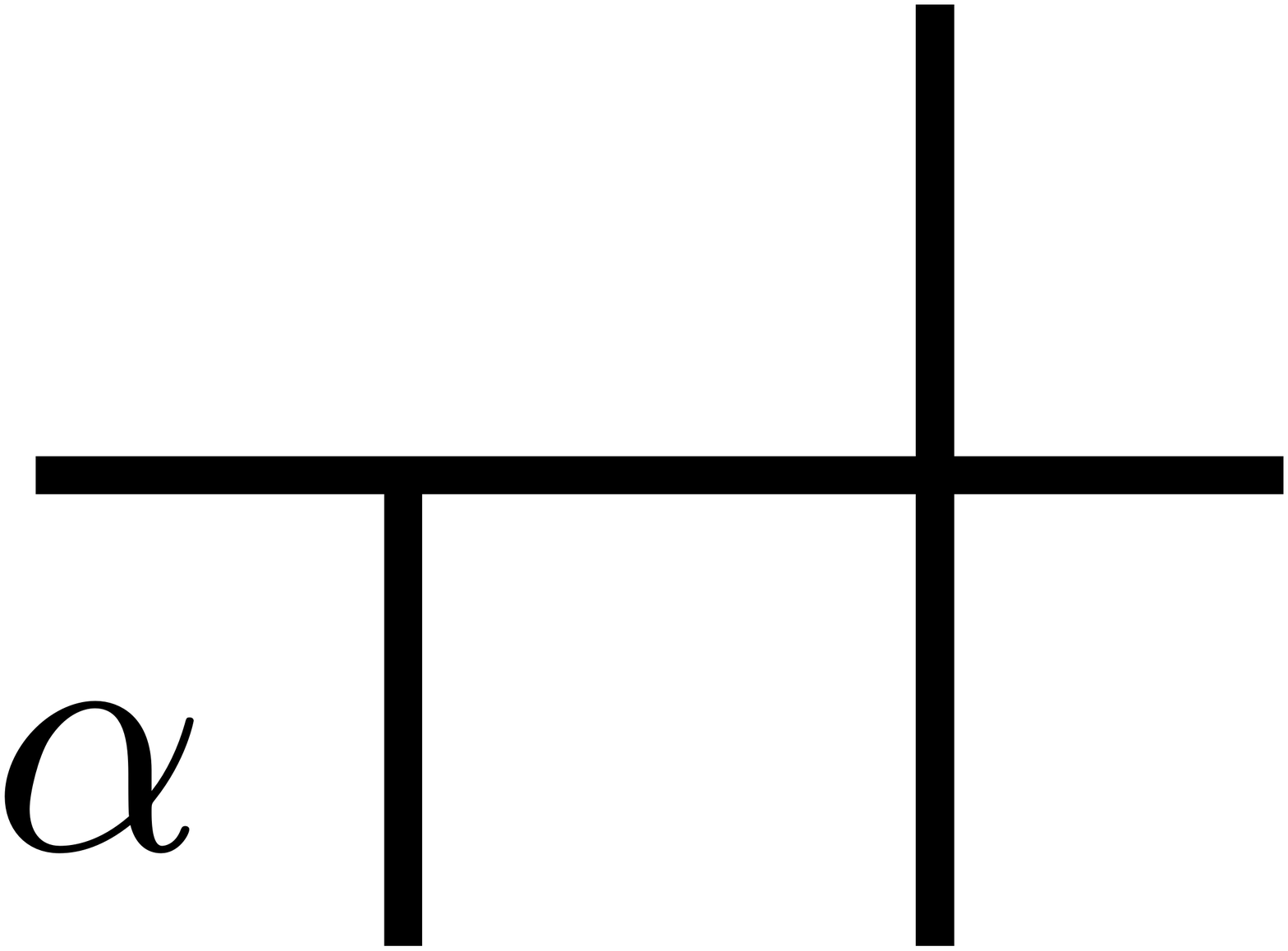}}\,\,\,+\,\,\, \frac{1}{2} \underbrace{v_3^3 \sum_{\alpha,\beta}   G(\alpha) G(\beta)}_{\includegraphics[scale=0.06]{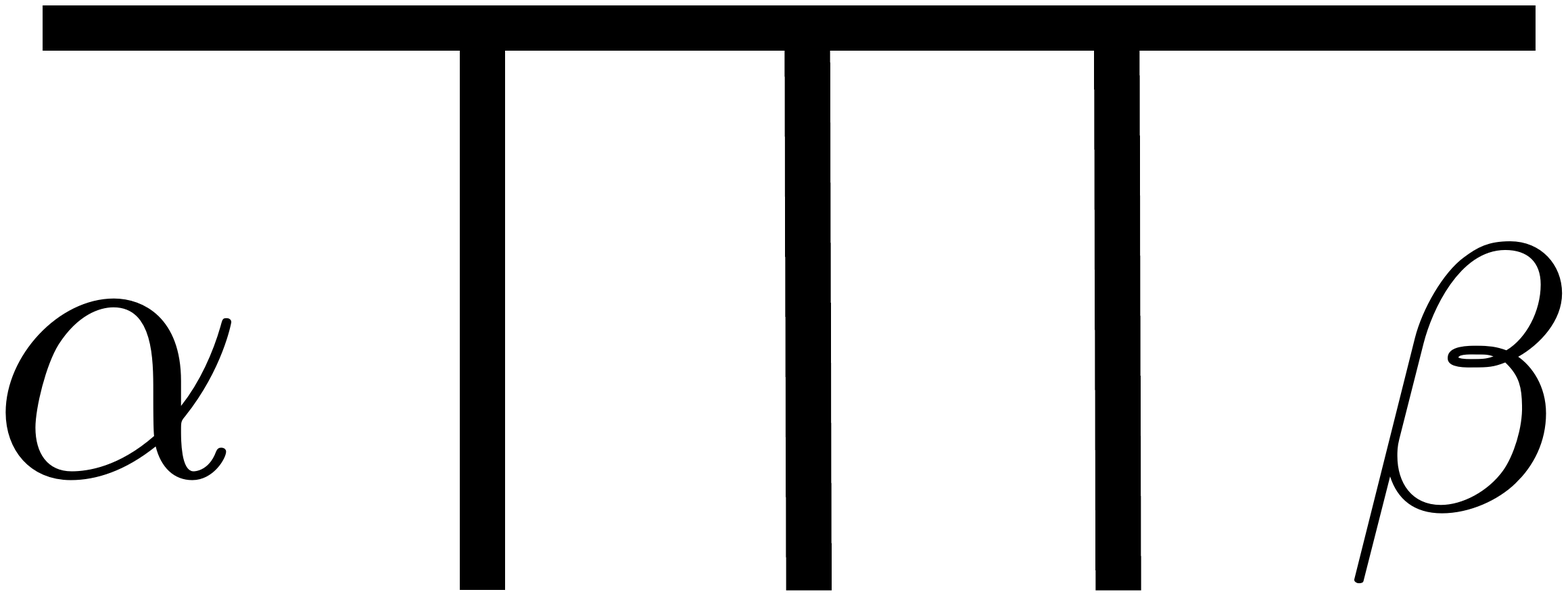}}\,\,\,+\!\!\!\!\!\!\underbrace{{\color{white}\frac{1}{2}\sum_\alpha\!\!\!\!\!\!\!\!\!\!\!\!\!\!}v_5}_{\includegraphics[scale=0.06]{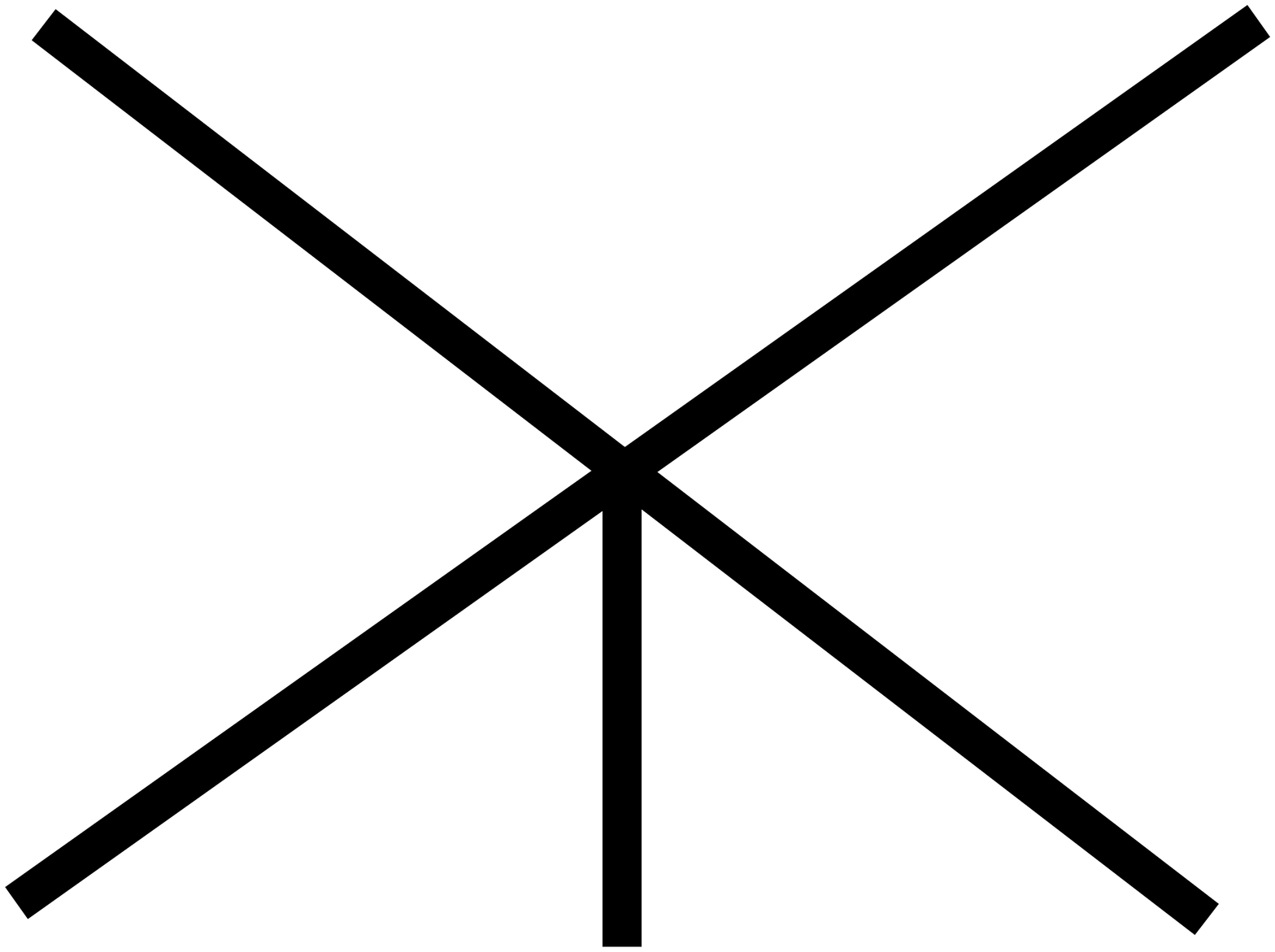}} \label{seed}\,,
\end{equation}
where $\alpha, \beta$ run over disjoint two-element subsets of the set of external particles $\{1,\ldots,5\}$ and where the (rescaled) propagator takes the form
\begin{equation}
G(\alpha)= \frac{1}{(\sum\limits_{j\in\alpha}a_j)(\sum\limits_{k\in\alpha}a^{-1}_k)-1}\,.
\end{equation}
Total energy-momentum conservation reads $\sum_{j=1}^5 a_j = \sum_{j=1}^5 a^{-1}_j = 0$. Rather remarkably, on the support of these constraints \textit{and} for $v_4=3v_3^2$ the first two terms in (\ref{seed}) sum to a constant and can thus be cancelled by appropriately tuning the last term.

Having cancelled three-particle production by setting $v_4=3v_3^2$, we can now move on to $\mathcal{M}_{2\to 4}$ where we get 
\begin{equation}
\centering
\includegraphics[scale=0.55, valign=c, trim={0 13cm 0 0.5cm}]{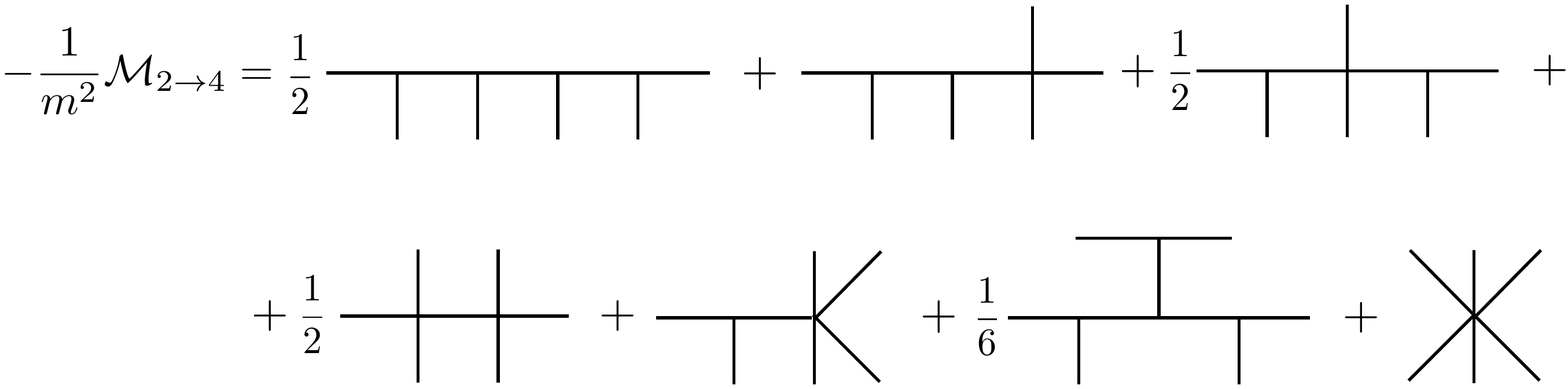} \,.
\label{eq:2to4Intro}
\end{equation}
Again, we find that on the support of energy-momentum conservation, the first six terms sum to a constant and can thus be cancelled by an appropriate choice of $v_6$ in the last term. It is possible to proceed in this way and find that one can always cancel $\mathcal{M}_{2\to n-2}$ by appropriately fixing $v_n$. After some tedious calculations, this leads to
\begin{equation}
\mathcal{L}_\text{interaction}=-m^2 \left(\frac{v_3}{3!}\phi^3+\frac{3v_3^2}{4!}\phi^4+\frac{5v_3^3}{5!}\phi^5+\frac{11v_3^4}{6!}\phi^6+\frac{21v_3^5}{7!}\phi^7+\dots\right) \label{data}
\end{equation}
At this point we could try to guess the result. Instead, we would like to proceed more systematically.

\begin{figure}[t]
\centering
\includegraphics[scale=0.4]{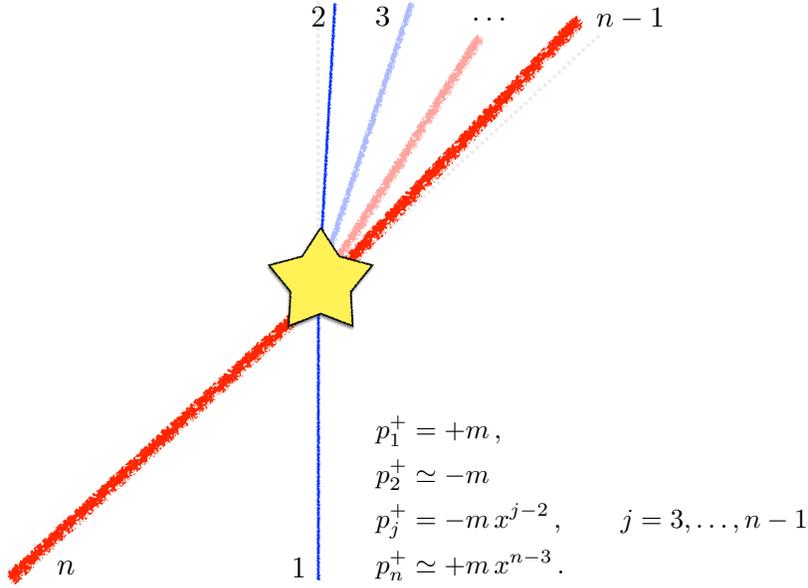}
\caption{The precise expression for the momenta of particles $2$ and $n$ are given by solving energy-momentum conservation~$\sum p_j^+ = \sum 1/p_j^+ = 0$. In the multi-Regge limit with $x\gg 1$ we have a highly energetic particle hitting a particle at rest producing a particle which is almost at rest plus a shower of very energetic particles. (More precisely, one finds the momenta $p_{n}^+ = - p_{n-1}^+-\dots-p^+_{3}+O(x^{-1}) \simeq m x^{n-3}$ and~$p_2^-=-p_1^--p_3^-  - \dots - p_{n-2}^- +O(x^{2-n}) \simeq -m$.) } \label{LimitFig}
\end{figure}

Let us for the time being operate under the assumption that particle production can be exactly cancelled and let us try to fix the coupling constants that guarantee it. We will defer the proof of the possibility of complete cancellation to Section \ref{sec:miracle}. The key idea that allows us to fix the couplings uniquely is to introduce a convenient multi-Regge limit where one incoming particle is at rest, with light-cone momenta $p_1=m(1,1)$, and $n-3$ outgoing particles are very energetic with
\begin{equation}
p_j= (p_{j}^+,p_{j}^-) = -m(x^{j-2},1/x^{j-2}) \, , \qquad j=3,\dots,n-1\,,  \label{momentaX}
\end{equation}
where $x$ is taken to be very large and positive. The momenta $p_2$ and $p_n$ of the remaining two particles are fixed by momentum conservation. We find that (without loss of generality) particle 2 is outgoing and almost at rest while particle $n$ is incoming and highly energetic. The configuration is illustrated in Figure \ref{LimitFig}.

\begin{figure}[t]
\centering
\includegraphics[scale=0.55]{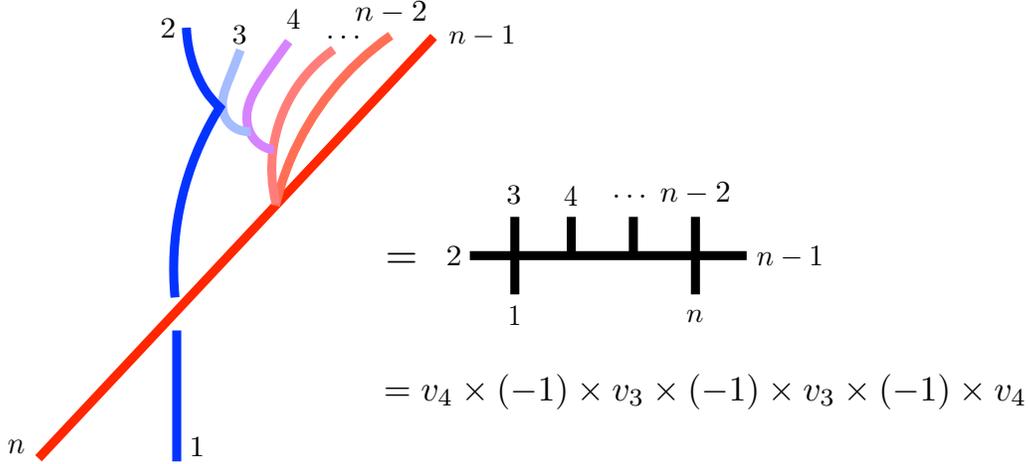}
\vspace{-5cm}\caption{The only surviving diagrams in the multi-Regge limit are one-dimensional chains with particles ordered along the chain. They evaluate to the product of involved vertices and $(-1)$ per propagator.} \label{ExampleFig}
\end{figure}

At tree level, any propagator separates a subset $\alpha\subset\{1,\ldots,n\}$ of the external particles from its complement. Most such subsets make highly energetic jets and thus vanishingly small propagators. The only propagators which survive in the limit $x\rightarrow\infty$ are the ones where particles $\{1,2,\dots,j-1\}$ are on one side and particles $\{j,j+1,\dots,n-1,n\}$ on the other so that the momentum transfer is small. Specifically,
\begin{equation} 
\!\!\! \lim_{x\to \infty} G(\alpha) = \left\{ \begin{array}{ll}
-1 &\text{if } \alpha=\{j,j+1,\ldots,n\} \qquad (\text{or equivalently } \alpha=\{1,2,\ldots,j-1\})\\ \\
0 & \text{otherwise}\,.
\end{array} \right.  \label{Outcome}
\end{equation}
Hence, the only surviving tree-level Feynman graphs are one-dimensional chains with all particles ordered. Figure \ref{ExampleFig} shows an example of a surviving Feynman diagram. For example, the $2\to 4$ amplitude (\ref{eq:2to4Intro}) immediately simplifies to $-\mathcal{M}_{2\to 4}/m^2 = -v_3^4+2 v_3^2v_4 + v_3^2v_4 - v_4^2 + 2v_3v_5 + 0 + v_6 \,.$

\begin{figure}[h]
	\begin{center}
		\def\svgwidth{16cm}
		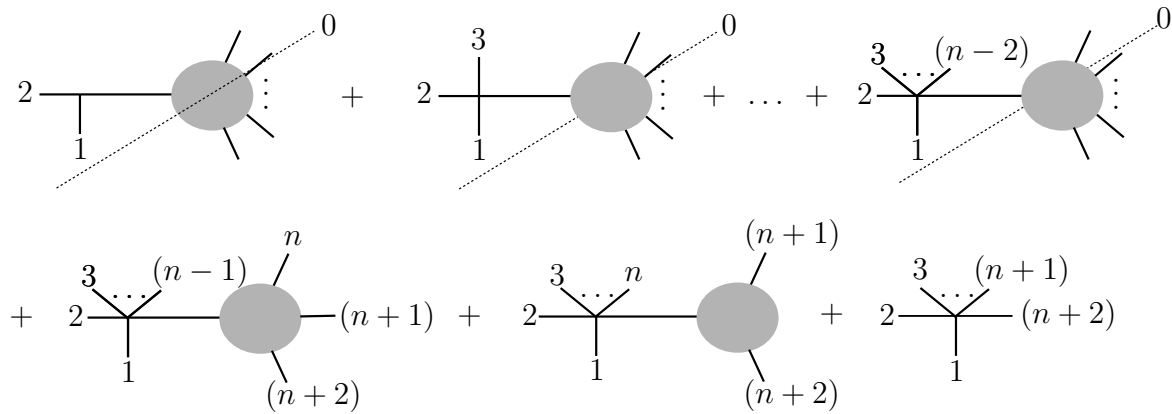
		\caption{\label{fig:SumRecurIntro}We organize all ordered diagrams so that particle $1$ is always on the left. The sum of diagrams where particle 1 is attached to a $k$-point vertex with $k<n$ evaluates to zero since it is attached to a total amplitude with $n-k+4>4$ external particles. As a result, only the three contributions in the last row survive.}
	\end{center}
\end{figure}

It is now easy to find the general Lagrangian by induction. We consider~$\mathcal{M}_{2\to n}$ assuming $\mathcal{M}_{2\to 3}, \dots,\mathcal{M}_{2\to n-1}$ were already tuned to vanish by fixing the vertices up to~$v_{n+1}$. The amplitude $\mathcal{M}_{2\to n}$ in the multi-Regge limit is given by a sum of one-dimensional ordered chains. Particle $1$ must therefore be at an end-point of such chains and can interact through a vertex of any valency, as illustrated in Figure \ref{fig:SumRecurIntro}. The only surviving graphs are those where the vertex is an $n$-, $(n+1)$- or $(n+2)$-particle vertex since those are respectively dressed by $4$, $3$ and $2$ total particle amplitudes which are the only non-zero amplitudes (since $5,6,\dots,n+1$ were already constrained to vanish, by assumption). We thus find

\begin{equation}
-\frac{1}{m^2}\mathcal{M}_{2\to n}= \underbrace{v_{n+2}}_{\includegraphics[scale=0.06]{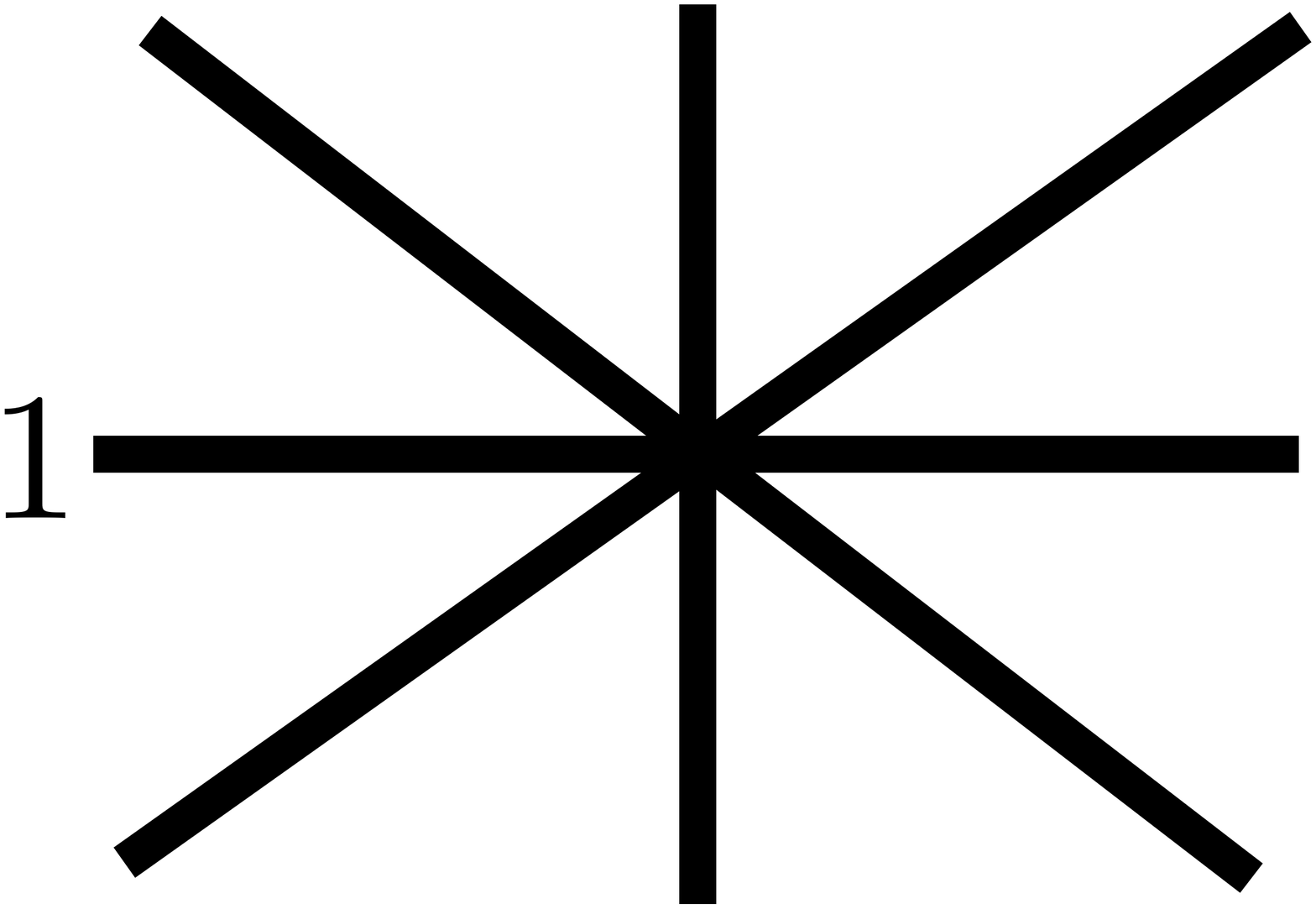}}+\,\,\,\underbrace{(-v_{n+1} v_3) }_{\includegraphics[scale=0.06]{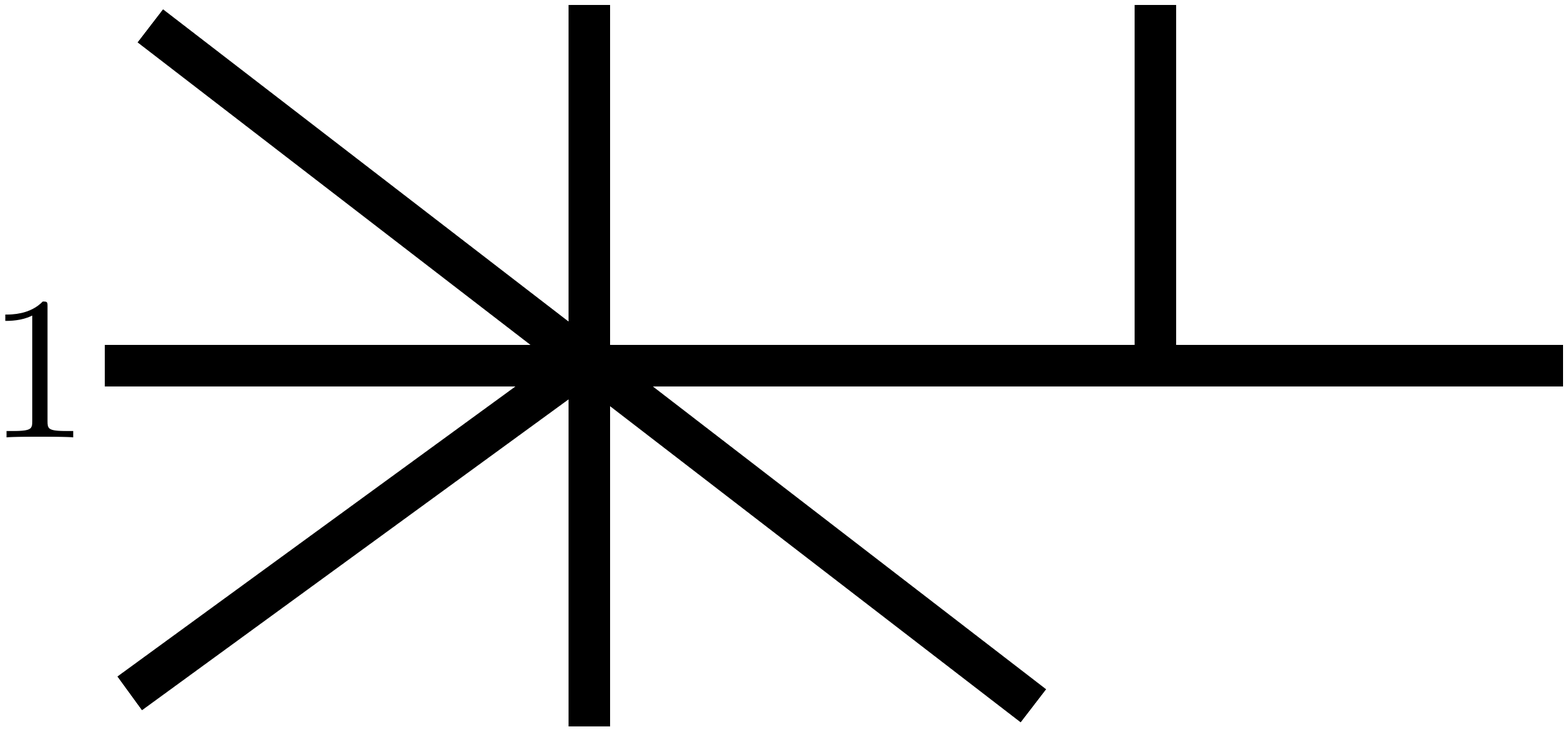}}\,\,\, +\underbrace{v_n \left(v_3^2-v_4\right)}_{\includegraphics[scale=0.06]{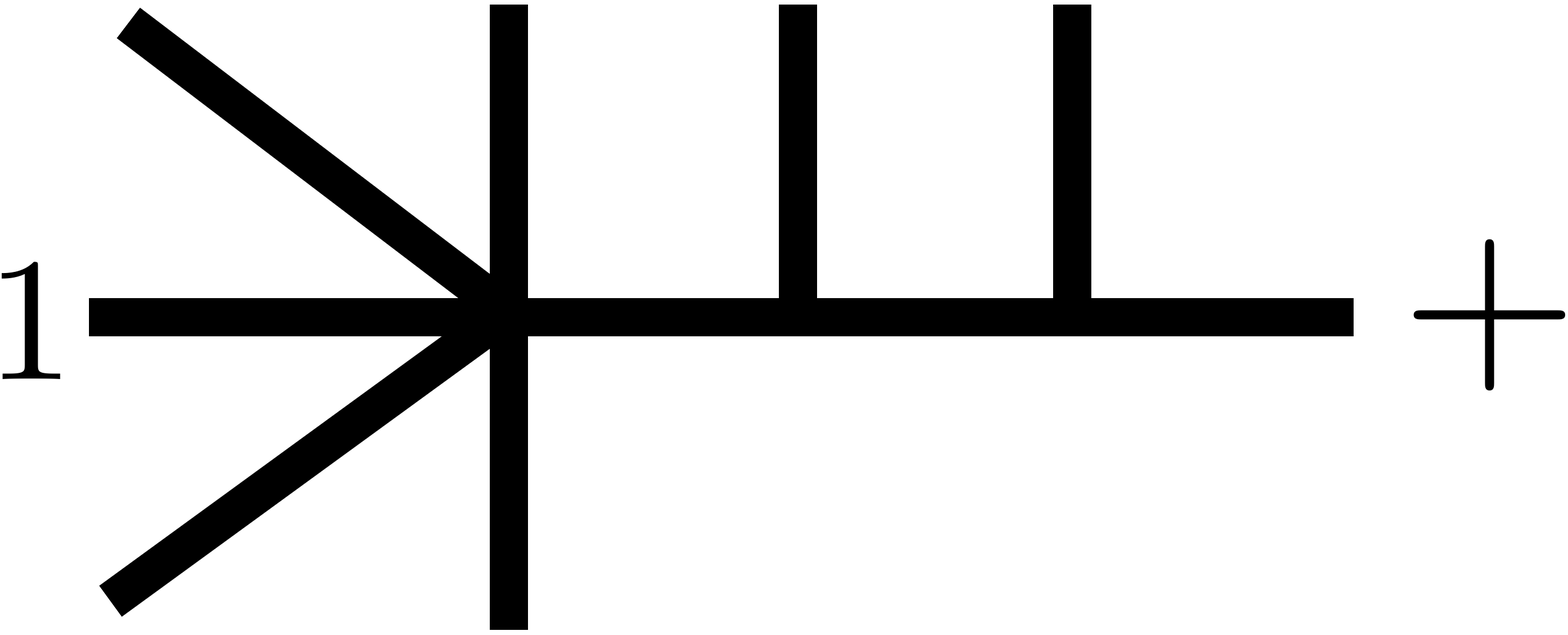}\includegraphics[scale=0.06]{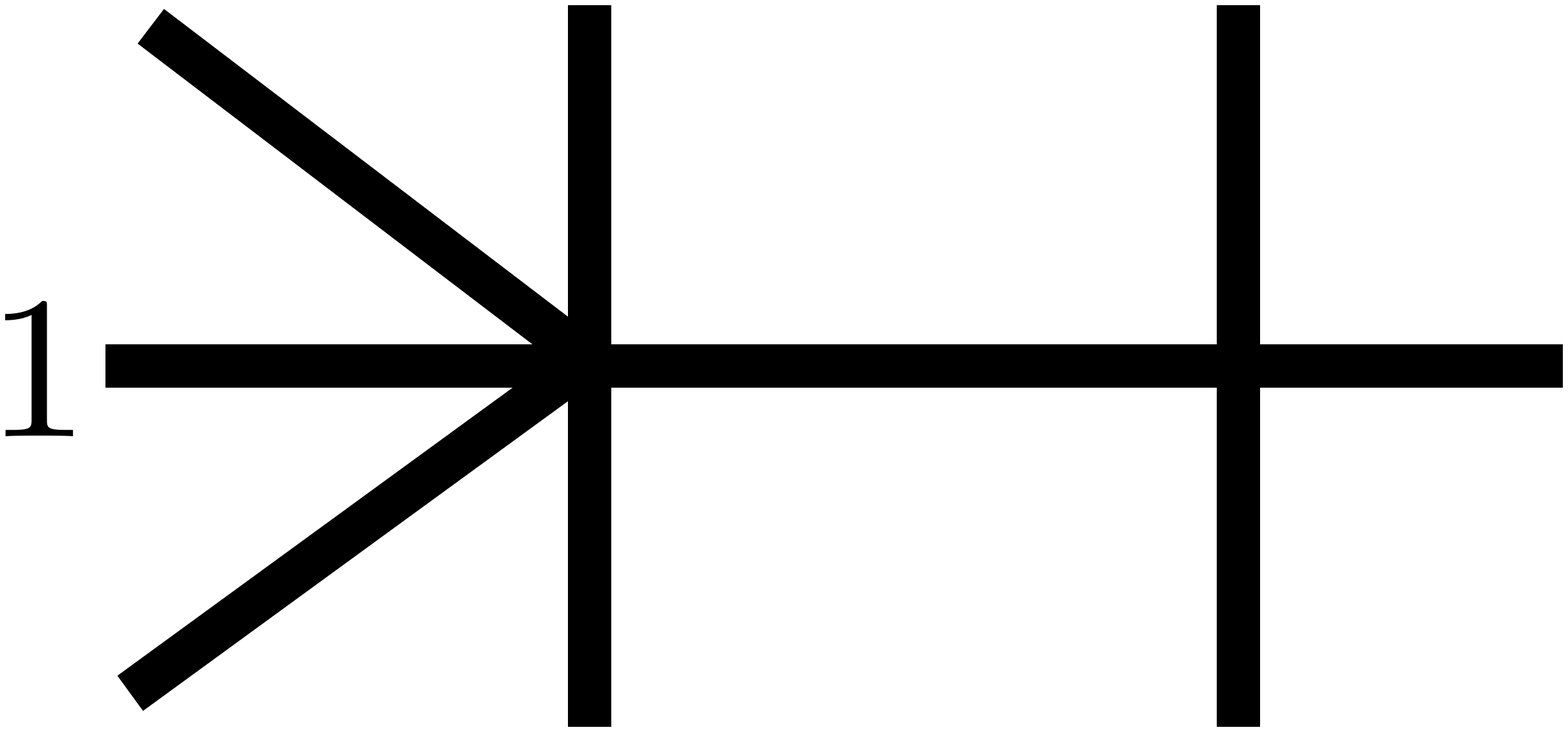}}\label{recursion}
\end{equation}
Recalling that $v_4=3v_3^2$ and requiring this amplitude to vanish, we obtain the desired recursion relation which one can readily solve, 
\begin{equation}
0= v_{n+2}-v_{n+1} v_3 - 2 v_n v_3^2 \qquad \Rightarrow \qquad v_n = \frac{2+(-2)^{n}}{6}\lambda^{n-2}  \,,  
\end{equation}
thus obtaining the famous Bullough-Dodd model,
\begin{equation}
\mathcal L_{BD} = \frac{1}{2} (\partial \phi)^2 - \frac{m^2}{6\lambda^2} \left[2e^{\lambda \phi}+ e^{-2\lambda \phi} - 3 \right]\, . 
\end{equation}
We see that this is the only theory with a single massive scalar particle, a cubic coupling, a perturbative expansion with no derivative couplings and no particle production at tree-level. It is a pleasure to check that the Taylor expansion of this potential does match with the painfully obtained data in (\ref{data}).

We can also study $\mathbb{Z}_2$-symmetric scalar theories, i.e. those where all odd-point interaction vertices vanish. In this case, we can set $v_3=0$ in (\ref{recursion}) to obtain the simpler $\mathbb{Z}_2$ recursion relation $v_{n+2}-v_n v_4=0$ leading to $v_{n} = \beta^{n-2}$ for $n$ even so that the potential resums to the sinh-Gordon theory\footnote{Or sine-Gordon if we take $\beta$ to be purely imaginary.}
%
\begin{equation}
\mathcal L_{sG} = \frac{1}{2} (\partial \phi)^2 - \frac{m^2}{\beta^2} \left[\cosh(\beta \phi) - 1 \right]\ .
\end{equation}

In Section \ref{sec:miracle} we tie up some loose ends of the above derivation. First, we discuss in more detail the seeds of the recursion relations, i.e. the remarkable identities which state that particular sums of diagrams arising in lower amplitudes actually add up to constants. Second, we explain in more detail why, once this is established, we are guaranteed to be able to cancel higher-particle production by suitably adjusting the higher-point interaction vertices. We can then pick any simplifying kinematics to find these couplings and the multi-Regge derivation we just described is a particularly convenient choice.  The method generalizes to other theories. Section \ref{sec:generalizations} contains a very preliminary start of such explorations. There, we comment on generalizations to a higher number of fields, theories with derivative interactions and theories with color orderings and make contact with Toda theories and non-linear sigma models. We were told that a supersymmetric analysis is to appear in \cite{brotherPaper} following similar techniques.

Can any of this can shed light on the very small particle production observed in the recent higher-dimensional S-matrix bootstrap explorations? Perhaps in a similar multi-Regge limit we can develop some intuition? Or perhaps, the better analogy is in terms of enhanced soft limits?
We should explore this further. 

\section{No particle production via analytic properties}
\label{sec:miracle}
In this section, we tie up the two loose ends of the discussion above. First, we discuss in more detail the seeds of the recursion relations, i.e. the remarkable identities which state that particular sums of diagrams arising in lower-point amplitudes actually add up to constants. Second, we explain in more detail why, once this is established, we are guaranteed to be able to cancel higher-particle production by suitably adjusting the remaining interaction vertices.

\subsection{General comments}
\label{sec:GenCom}
Let us treat all particles as incoming and parametrize the external momenta by $a_j$ so that $p_j = m(a_j,1/a_j)$ in light-cone coordinates. Since we want to cancel tree-level particle production in a generic kinematical configuration, we can set $\epsilon=0$ in the $i\epsilon$ prescription. This is because a nonzero $\epsilon$ can only introduce additional momentum-space delta-functions in the $\epsilon\to0$ limit (at tree level). We will denote the $n$-point scattering amplitude by $\mathcal{M}_n$. Firstly, we would like to comment on the complex-analytic properties of $\mathcal{M}_n$. The Feynman-diagrammatic prescription gives $\mathcal{M}_n$ as a rational function of all $a_j$, $j\in\{1,\ldots,n\}$. However, the $a_j$s satisfy a pair of algebraic constraints corresponding to momentum conservation $\sum_{j=1}^n a_j = \sum_{j=1}^n a^{-1}_j = 0$. We can solve these constraints to find, say $a_{n-1}, a_{n}$ in terms of $a_1,\,\ldots,a_{n-2}$. The solution contains square roots, reflecting the fact that the constraints are symmetric under the transformation $a_{n-1}\leftrightarrow a_{n}$. Fortunately, $\mathcal{M}_n$ is also symmetric under this transformation. This guarantees that after substituting for $a_{n-1}, a_{n}$ in terms of $a_1,\,\ldots a_{n-2}$, all square roots drop out and $\mathcal{M}_n$ becomes a symmetric rational function of the independent variables $a_1,\,\ldots,a_{n-2}$. By permutation symmetry, we can think of $\mathcal{M}_n$ as a rational function of any $(n-2)$-element subset of the $a_j$s.

Another useful property of $\mathcal{M}_n$ is that it is left invariant under the simultaneous rescaling $a_j\mapsto \lambda a_j$. Now, imagine we can demonstrate that $\mathcal{M}_n$ has no poles as a function of any of the $a_j$s. Since it is a rational function, it must be a polynomial. The invariance under the simultaneous rescaling then shows that it must in fact be a constant. Therefore, to demonstrate the constancy of a given amplitude, it is sufficient to show it has no poles.

Our proof that particle production can be cancelled at tree-level in the sine-Gordon and Bullough-Dodd theories proceeds by induction on the number of external particles. First, we will analyze the base cases $\mathcal{M}_5$ and $\mathcal{M}_6$, and then move on to proving the induction step.

\subsection{Base case for sine-Gordon}
\label{sec:SGSingularityStruct}

Since $\mathcal{M}_5=0$ in the sine-Gordon theory by the $\mathbb{Z}_2$ symmetry, it is enough to analyze $\mathcal{M}_6$. We want to show that on the support of momentum conservation, the function
\begin{equation}\label{goal}
\sum_{\sigma}\!\!\!\!\underbrace{G(\sigma)}_{{\includegraphics[scale=0.07]{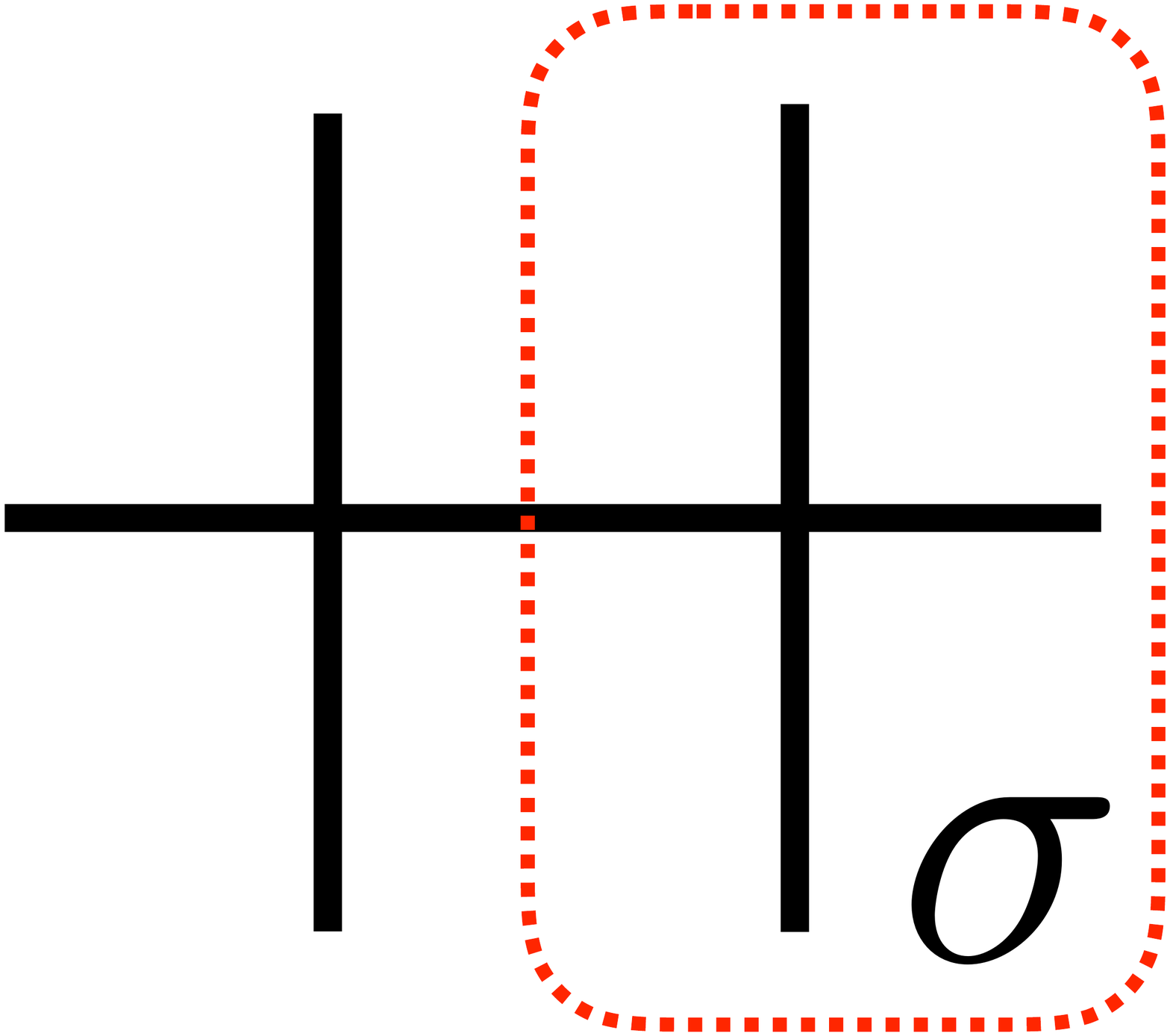}}} = \sum_{\sigma} \frac{a_{\sigma(1)}a_{\sigma(2)}a_{\sigma(3)}}{(a_{\sigma(1)}+a_{\sigma(2)})(a_{\sigma(1)}+a_{\sigma(3)})(a_{\sigma(2)}+a_{\sigma(3)})}
\end{equation}
has no poles as a function of the $a_j$s, where the sum runs over all three-element subsets of $\{1,\ldots,6\}$. Poles may only occur when $a_j\to - a_k$. Thanks to the symmetry under arbitrary permutations of the external particles, it is enough to look at the one when $a_5 \to -a_6$. When this happens, particles $5$ and $6$ annihilate each other and disappear from the momentum conservation constraints which thus become $ \sum_{j=1}^4 a_j = \sum_{j=1}^4 1/a_j =0 $. These still admit several two-parameter branches of solutions. Thanks to symmetry, we can pick one of the branches, say $a_1=-a_2$ and $a_3=-a_4$. We ended up with three pairs of particles all annihilating at the same time. The terms of \eqref{goal} singular in this limit must have any of the pairs $(12)$, $(34)$ or $(56)$ either inside $\sigma$ or in its complement. There are six such terms. For example for $\sigma=\{1,2,3\}$, we find
\begin{equation}
\res_{a_1\to-a_2} \frac{a_1 a_2 a_3}{(a_1+a_2)(a_2+a_3)(a_3+a_1)} = \frac{a^2_2\, a_3}{a_2^2-a_3^2}\ . \label{ResProp3}
\end{equation}
But this term cancels with the term where $a_3$ is replaced by $a_4$ since $a_3$ is approaching $-a_4$ in this limit. All other terms cancel in the same pairwise fashion thus showing that \eqref{goal} is a constant.
 
\subsection{Base cases for Bullough-Dodd}
\label{sec:BaseBD}
The base cases of our argument for the BD model consist of showing the constancy of $\mathcal{M}_5$ and $\mathcal{M}_6$. For $\mathcal{M}_5$, we need to demonstrate that on the support of momentum conservation, the quantity 
\begin{equation}
\begin{aligned}
&\sum_{\sigma} \frac{a_{\sigma(1)}a_{\sigma(2)}a_{\sigma(3)}}{(a_{\sigma(1)}+a_{\sigma(2)})(a_{\sigma(1)}+a_{\sigma(3)})(a_{\sigma(2)}+a_{\sigma(3)})}\times\\
&\underbrace{\times \left(1+ \frac{\lambda}{2} \sum_{\mu} \frac{a_{\mu(1)}a_{\mu(2)}a_{\mu(3)}}{(a_{\mu(1)}+a_{\mu(2)})(a_{\mu(1)}+a_{\mu(3)})(a_{\mu(2)}+a_{\mu(3)})} \right) }_{{\includegraphics[scale=0.2, trim={0 10cm 0 0}]{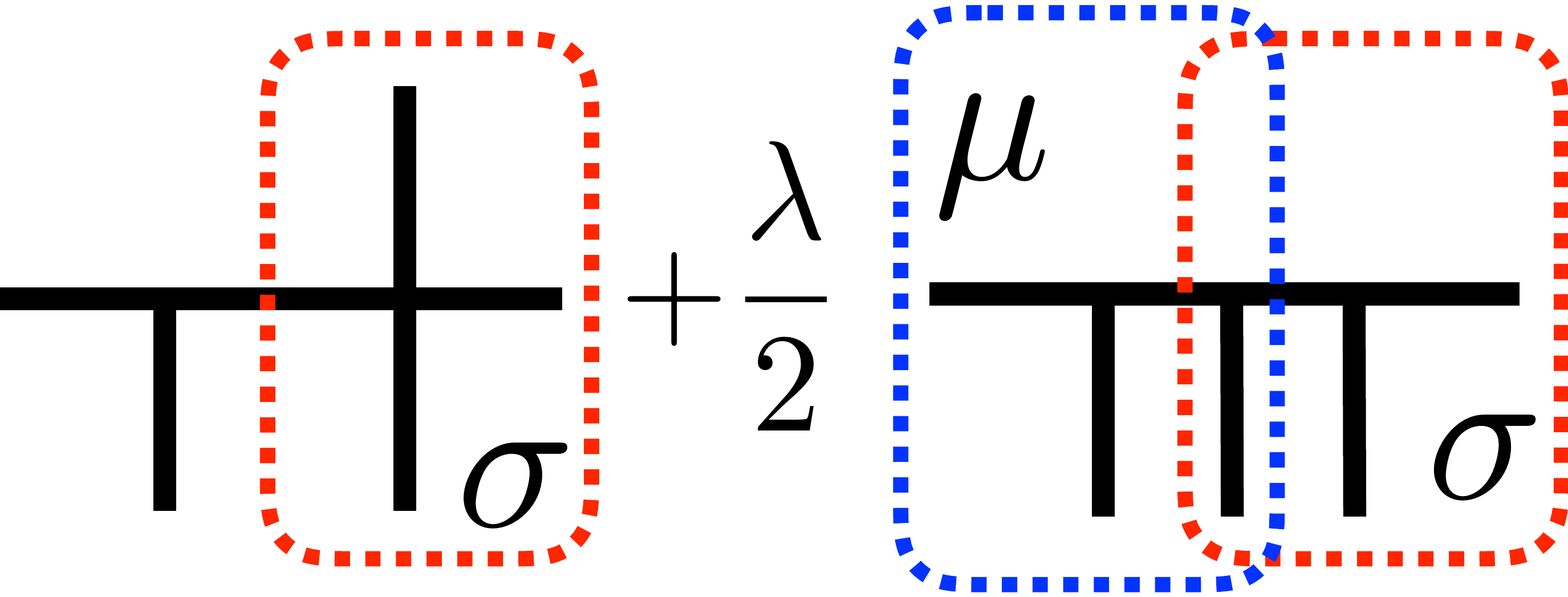}}}
\end{aligned}
\end{equation}
has no poles for properly tuned $\lambda$. $\sigma$ runs over all three-element subsets of $\{1,\ldots,5\}$ and $\mu$ runs over three-elements subsets of $\{1,\ldots,5\}$ sharing exactly one element with $\sigma$. Again, it suffices to analyze the pole that occurs as $a_1\to -a_2$. The pole could come from the $\mu$ propagator or the $\sigma$ propagator. Hence, using (\ref{ResProp3}) we get the residue
\begin{eqnarray} \nonumber
&&\!\!\!\!\!\!\!\!\!\!\!\!\!\!\!\sum_{j=3,4,5} \frac{a^2_2\, a_j}{a_2^2-a_j^2} \Big(1+2  \frac{\lambda}{2} \underbrace{ \frac{a_3 a_4 a_5}{(a_3+a_4)(a_3+a_5)(a_4+a_5)}}_{=-1 \text{ for } a_1\to -a_2}+ 2  \frac{\lambda}{2}\underbrace{\sum_{  l=1,2}   \sum_{  k,r \neq j
}  \frac{a_l a_r a_k}{(a_l+a_r)(a_l+a_k)(a_r+a_k)}}_{=-\frac{2 a_2^2 a_r a_k}{\left(a_2^2-a_r^2\right) \left(a_2^2-a_k^2\right)} \text{ for } a_1\to -a_2}\Big) \\
&&\qquad \qquad=\frac{a_2^2}{\prod_{k=3}^5(a_2^2-a_k^2)}\Big((1-\lambda)\!\!\!\!\!\!\!\!\!\!\!\!\!\!\!\! \underbrace{\sum_{k} a_k\prod_{l\neq k}(a_2^2-a_l^2)}_{ {\color{red} a_2^4} (\sum\limits_k a_j)-2 {\color{blue} a_2^2} (\sum\limits_{k\neq l} a_k a_l^2)+{\color{red}a_2^0} a_3 a_4 a_5 (\sum\limits_{k\neq l} a_k a_l)   }\!\!\!\!\!\!\!\!\!\!\!\!\!\!\!\!-\,3 \times 2 \lambda {\color{blue}a_2^2} a_3 a_4 a_5 \Big)
\end{eqnarray}
Now, because of momentum conservation we have $\sum_{k=3}^5 a_j = \sum_{k\neq j}^5 a_k a_j = 0$ so only the terms quadratic in $a_2$ survive. Finally, multiplying both momentum conservation constraints yields $0=(a_3+a_4+a_5)(a_4 a_5 + a_3 a_5+a_3 a_4)$ yields $2 \sum_{k\neq j} a_k a_l^2=-3a_3a_4a_5$ so this remaining quadractic term is simply equal to ${\color{blue}a_2^2} a_3 a_4 a_5 (3(1-\lambda)-6\lambda)$ and thus vanishes for $\lambda=1/3$. We have thus shown that provided $v_4=3v_3^2$, $\mathcal{M}_5$ has no poles and thus is a constant.

Provided the constant $\mathcal{M}_5$ is cancelled by an appropriately chosen $v_5$, it is a simple matter to also demonstrate the constancy of $\mathcal{M}_6$, which we leave as an exercise to the reader.

\subsection{The induction step}
The purpose of this subsection is to prove the step of our induction. Specifically, we would like to show that if $\mathcal{M}_{j}$ vanishes for all $j\in\{5,\ldots, n\}$ and $n\geq 6$, then the $n+1$-point coupling can be chosen so that also the $\mathcal{M}_{n+1}$ amplitude vanishes.

Let us assume that $\mathcal{M}_j$ vanishes for $j=5,\ldots,n$ with $n\geq 6$ and study $\mathcal{M}_{n+1}$ as a function of the complex variable $a_1$ with $a_2,\ldots,a_{n-1}$ generic. The only allowed singularities correspond to a single propagator going on-shell. This is because a given internal propagator separates a tree amplitude into two subamplitudes. When such propagator goes on-shell, we obtain two on-shell amplitudes. An on-shell amplitude is non-singular for generic momenta, implying no additional singularities occur for generic external momenta with a single internal propagator going on-shell.

Take the propagator to separate the external legs into subsets $A$, $\{1,\ldots n+1\}\backslash A$ and denote $p_0=\sum_{i\in A}p_i$. The residue at the pole is propotional to a product of lower-point on-shell amplitudes
\begin{equation}
\text{Res}_{p_0^2\rightarrow 0}\mathcal{M}_{n+1}\sim \mathcal{M}_{k+1}(a_0,a_i) \mathcal{M}_{n-k+2}(-a_0,a_j)\,,
\label{eq:factorization}
\end{equation}
where $k=|A|$, $i\in A$ and $j\in \{1,\ldots n+1\}\backslash A$. Since $2\leq k\leq n-1$, one of the two on-shell subamplitudes vanishes by the induction hypothesis. Indeed, the first factor on the RHS of \eqref{eq:factorization} vanishes if $k\geq 4$, and the second factor certainly vanishes in the remaining cases $k=2,3$. We conclude $\mathcal{M}_{n+1}$ has no poles as a function of $a_1$. 

The same argument applies to $\mathcal{M}_{n+1}$ as a function of the remaining variables $a_2,\,\ldots,a_{n-1}$ and we conclude $\mathcal{M}_{n+1}(a_1,\ldots,a_{n-1})$ is a polynomial. The Feynman-diagrammatic definition implies that $\mathcal{M}_{n+1}$ is invariant under the simultaneous rescaling $a_i\mapsto \lambda a_i$, so that the polynomial is in fact a constant. The coupling $v_{n+1}$ can now be chosen to cancel this constant, which completes the proof of the induction step.

\section{Generalizations}
\label{sec:generalizations}
\subsection{Multiple fields}
In this subsection, we will generalize parts of the above analysis to the case of multiple scalar fields. We consider the most general Lagrangian for a two-dimensional theory with $N$ real scalar fields $\phi_\alpha$, $\alpha=1,\ldots,N$ with non-derivative interactions:
\begin{equation}
\mathcal{L} = \frac{1}{2}(\partial_\mu\phi_\alpha)(\partial^\mu\phi_\alpha)-\frac{1}{2}m_\alpha^2\phi_\alpha^2 - \sum\limits_{n=3}^{\infty}\frac{v^{\alpha_1\ldots \alpha_n}_n}{n!}\phi_{\alpha_1}\ldots\phi_{\alpha_n}\,, \label{LagMulti}
\end{equation}
where repeated field indices are summed over from 1 to $N$ and $v^{\alpha_1\ldots \alpha_n}_n$ is a totally symmetric tensor of rank $n$. As before, we would like to constrain $v_n^{\alpha_1\ldots \alpha_n}$ by imposing the theory has no particle production at tree-level. Since the external particles can be arbitrary, this requirement clearly fixes all $n$-point vertices for $n\geq 5$ in terms of the cubic and quartic couplings. We would like to find the recursion relation on the couplings analogous to \eqref{recursion}, using the obvious analogue of the multi-Regge limit \eqref{momentaX}. We consider the scattering of $n$ particles of species $\alpha_1,\ldots,\alpha_n$ and parametrize their light-cone momenta using the variables $z_j$ as $p_j = m_j(z_j,1/z_j)$. In the multi-Regge limit, we take
\begin{equation}
z_1 = 1\textrm{ and }z_j = -x^{j-2}\,\textrm{ for }j=3,\ldots,n-1
\end{equation}
with $x\rightarrow\infty$. We use momentum conservation to solve for $z_2$ and $z_n$. On one of the two branches of solutions, we find
\begin{equation}
z_2 = -\frac{m_2}{m_1} + O(x^{-1})\textrm{ and }z_n = \frac{m_{n-1}}{m_n}x^{n-3} + \frac{m_{n-2}}{m_n}x^{n-4} + \ldots +\frac{m_3}{m_n}x + O(x^0)\,.
\end{equation}
Again, the only diagrams which survive the $x\rightarrow\infty$ limit are linear chains. The rest of the derivation of the recursion relation is identical to the single field case, except for the need to sum over particle species in internal propagators. To write the recursion relation, it is first convenient to define
\begin{equation}
\tilde{v}_n^{\alpha_1\ldots\alpha_n} = \frac{v_n^{\alpha_1\ldots\alpha_n}}{m_{\alpha_1}\ldots m_{\alpha_n}}\,.
\end{equation}
The recursion then reads
\begin{equation}
\tilde{v}_n^{\alpha_1\ldots\alpha_n} = \tilde{v}_{n-1}^{\alpha_1\ldots\alpha_{n-2}\beta}\tilde{v}_3^{\beta \alpha_{n-1}\alpha_{n}} + \tilde{v}_{n-2}^{\alpha_1\ldots\alpha_{n-3}\beta}\left(\tilde{v}_4^{\beta\alpha_{n-2}\alpha_{n-1}\alpha_{n}} - \tilde{v}_3^{\beta\alpha_{n-2}\gamma}\tilde{v}_3^{\gamma\alpha_{n-1}\alpha_{n}}\right)\,,
\label{eq:recursionN}
\end{equation}
where repeated indices $\beta,\gamma$ are summed over. The recursion relation determines the $\tilde{v}_n^{\alpha_1\ldots\alpha_n}$ for $n\geq5$ in terms of $\tilde{v}_3^{\alpha\beta\gamma}$ and $\tilde{v}_4^{\alpha\beta\gamma\delta}$. Moreover, it turns out it imposes non-trivial constraints on $\tilde{v}_3^{\alpha\beta\gamma}$ and $\tilde{v}_4^{\alpha\beta\gamma\delta}$ too since the the right-hand side must be invariant under re-ordering the $\alpha$s since the left-hand side is fully symmetric. 
For example, when $N=2$, it allows us to fix the quartic couplings in terms of the cubic couplings as follows\footnote{Some of these equations can be given a nice physical meaning. For instance, consider an integrable theory with an arbitrary number of particles but where $m_2\neq m_1$. Then the inelastic amplitude $11\to 12$ should vanish and that amplitude is of course given at tree level by $v_{1112}+\sum_{x=1}^N v_{11x} v_{12x} (\frac{1}{s-m_x^2}+\frac{1}{t-m_x^2}+\frac{1}{u-m_x^2} )$. Vanishing of this component yields many constraints. One which is quite obvious is found at high energies when $s \to \infty$, then we have $t \to -\infty$ and $u\to 0$ so that we get simply $0=v_{1112}+\sum_{x=1}^N v_{11x} v_{12x}/(-m_x^2) $ which reduces to the second equation in (\ref{fiveEqs}) when $N=2$ and for $\hat v_4=0$. The third and fourth equation there have similar interpretations. 
}
\begin{equation}
\begin{aligned}
\tilde{v}_4^{1111} &= \hat{v}_4^{1111}+\tilde{v}_3^{111}\tilde{v}_3^{111} - \tilde{v}_3^{111}\tilde{v}_3^{122} + 2 \tilde{v}_3^{112}\tilde{v}_3^{112} + \tilde{v}_3^{122}\tilde{v}_3^{122}-\tilde{v}_3^{112}\tilde{v}_3^{222}\\
\tilde{v}_4^{1112} &=\hat{v}_4^{1112}+\tilde{v}_3^{111}\tilde{v}_3^{112}+\tilde{v}_3^{112}\tilde{v}_3^{122}\\
\tilde{v}_4^{1122} &=\hat{v}_4^{1122}+\tilde{v}_3^{112}\tilde{v}_3^{112}+\tilde{v}_3^{122}\tilde{v}_3^{122}\\
\tilde{v}_4^{1222} &=\hat{v}_4^{1222}+\tilde{v}_3^{112}\tilde{v}_3^{122}+\tilde{v}_3^{122}\tilde{v}_3^{222}\\
\tilde{v}_4^{2222} &=\hat{v}_4^{2222}+\tilde{v}_3^{112}\tilde{v}_3^{112}-\tilde{v}_3^{111}\tilde{v}_3^{122}+2\tilde{v}_3^{122}\tilde{v}_3^{122}-\tilde{v}_3^{112}\tilde{v}_3^{222}+\tilde{v}_3^{222}\tilde{v}_3^{222}\,,
\end{aligned} \label{fiveEqs}
\end{equation}
where $\hat{v}_4^{\alpha\beta\gamma\delta}$ is a solution of the following linear homogenous problem
\begin{equation}
\begin{pmatrix}
 \tilde{v}_{3}^{112} & \tilde{v}_{3}^{122}-\tilde{v}_{3}^{111} & -\tilde{v}_{3}^{112} & 0 & 0 \\
 -\tilde{v}_{3}^{122} & 2 \tilde{v}_{3}^{112}-\tilde{v}_{3}^{222} & 2 \tilde{v}_{3}^{122}-\tilde{v}_{3}^{111} & -\tilde{v}_{3}^{112} & 0 \\
 \tilde{v}_{3}^{122} & \tilde{v}_{3}^{222}-\tilde{v}_{3}^{112} & -\tilde{v}_{3}^{122} & 0 & 0 \\
 0 & \tilde{v}_{3}^{122} & \tilde{v}_{3}^{222}-\tilde{v}_{3}^{112} & -\tilde{v}_{3}^{122} & 0 \\
 0 & -\tilde{v}_{3}^{122} & 2 \tilde{v}_{3}^{112}-\tilde{v}_{3}^{222} & 2 \tilde{v}_{3}^{122}-\tilde{v}_{3}^{111} & -\tilde{v}_{3}^{112} \\
 0 & 0 & \tilde{v}_{3}^{122} & \tilde{v}_{3}^{222}-\tilde{v}_{3}^{112} & -\tilde{v}_{3}^{122} \\
\end{pmatrix} \cdot
\begin{pmatrix}
\hat{v}_4^{1111} \\ \hat{v}_4^{1112} \\ \hat{v}_4^{1122} \\ \hat{v}_4^{1222} \\ \hat{v}_4^{2222}
\end{pmatrix} =0\,.
\label{eq:mx}
\end{equation}
For generic values of the cubic vertices, the matrix has rank five, and therefore the only solution is $\hat{v}_4^{\alpha\beta\gamma\delta} = 0$. However, there are special values of the cubic vertices where the matrix degenerates and a nontrivial $\hat{v}_4^{\alpha\beta\gamma\delta}$ is allowed. It would be interesting to see whether there are further consistency constraints from the permutation symmetry of \eqref{eq:recursionN} for $n > 5$, and understand the space of solutions for $N>2$.

The recursion relation \eqref{eq:recursionN} is a necessary condition to have no tree-level particle production. In order to find sufficient conditions, we need to establish that the seeds of the recursion work by repeating the analysis of sections \ref{sec:SGSingularityStruct} and \ref{sec:BaseBD}. We performed this analysis for the case $N=2$, assuming the cubic vertices are such that the matrix in \eqref{eq:mx} has full rank, so we can use \eqref{fiveEqs} with $\hat{v}_4=0$. By imposing that the five-particle amplitude (which ought to vanish altogether) has no poles as a function of the external momenta, we find a set of six discrete solutions for the masses and cubic vertices. Only one of our solutions contains particles with equal masses and matches the $A_2$ affine Toda field theory \cite{ATFT-Mussardo,Braden:1991vz}, which has the following Lagrangian
\begin{equation} \label{eq:A2Toda}
\mathcal L_{A_2} = \frac{1}{2} (\partial_\mu \phi_\alpha\partial^\mu \phi_\alpha) -\frac{m^2}{3\beta^2} \left[e^{\sqrt{2}\beta\,\phi_1}+e^{\beta(\sqrt{3/2}\,\phi_2 - \phi_1/\sqrt{2})}+e^{-\beta(\sqrt{3/2}\,\phi_2 + \phi_1/\sqrt{2})} - 3 \right]\ .
\end{equation}
When the masses are distinct, our solutions correspond to the remaining five affine Toda field theories with two particles. In the notation of \cite{ATFT-Mussardo}, they have the following mass spectra:
\begin{equation}
\begin{aligned}
B_2 = C_2&:\quad m_2 = \sqrt{2}m_1\\
G_2&:\quad m_2 = \sqrt{3}m_1\\
A_3^{(2)} = D_3^{(2)}&:\quad m_2 = \sqrt{3}m_1\\\
A_4^{(2)}&:\quad m_2=\frac{1+\sqrt{5}}{2}m_1\\
D_4^{(3)}&:\quad m_2 = \sqrt{2+\sqrt{3}}m_1\,.
\end{aligned}
\end{equation}
We can now go back and check that indeed in all these cases, matrix in \eqref{eq:mx} has full rank. All cases of lower rank that we encountered correspond to a pair of decoupled sine-Gordon and Bullough-Dodd theories. sG+sG leads to rank zero, sG+BD to rank three and BD+BD also to rank three. If these decoupled cases are all there is, then we are done with the classification of theories with two fields and no particle production. Would be interesting to look for more exotic possibilities and explore the lower rank cases further.

Moving to $N>2$, we were able to check that the $B_4$ Toda, containing four particles, all of which have a distinct mass, satisfies the recursion relation \eqref{eq:recursionN}. The analysis of the seed problem for $N>2$ is beyond the scope of this work. It would be remarkable if one could use our algorithm to uncover theories with no classical particle production and Lagrangian of the form \eqref{LagMulti} which are not affine Toda field theories, thus plausibly discovering overlooked integrable field theories.

Finally, we can use the recursion \eqref{eq:recursionN} to demonstrate that there are no $O(N)$-symmetric theories of the type (\ref{LagMulti}) without particle production. Such theories would have vanishing cubic coupling and quartic couplings constrained by $O(N)$ symmetry to take the form~$v_4^{abcd}=A_1 \delta^{ab} \delta^{cd}+A_2 \delta^{ac} \delta^{bd}+A_3 \delta^{ad} \delta^{bc}$. Plugging this into the right-hand side of our recursion (\ref{eq:recursionN}) for $n=6$ and imposing that the right-hand side if fully symmetric leads to $A_j=0$ and hence to a trivial free theory where all couplings vanish. 

\subsection{Colour-ordered theories} 
\label{app:derIntOrd}
In this section we repeat the procedure done in the introduction for theories of Hermitian matrix valued massless fields interacting through two-derivative terms\footnote{Single scalar theories with two-derivative interactions $\mathcal L=\frac{1}{2}\text{Tr}(\partial _{\mu} \phi\partial ^{\mu} \phi)+\sum_{n=3}^{\infty} g_{n} \phi^{n-2} \partial_{\mu} \phi \partial ^{\mu}\phi$ are free theories in disguise since we can field refine the interactions away, $\mathcal L =\frac{1}{2}\text{Tr}(\partial\phi')^2$ with $\phi'=\phi+\tfrac{g_3}{2} \phi^2+\tfrac{2 g_4-g_3^2}{6}\phi^3+\dots$.} with ($g_{n,k}$ for $\lfloor{n/2}\rfloor<k$ are redundant due to cyclic invariance)
\begin{align}
\mathcal L(h)=\frac{1}{2}\text{Tr}(\partial _{\mu} h\partial ^{\mu} h)+\sum_{n=3}^{\infty}\sum_{k=1}^{\lfloor{n/2}\rfloor}g_{n,k} \underbrace{\text{Tr}(h^{k-1}\partial_{\mu} h h^{n-k-1}\partial ^{\mu}h)}_{{\includegraphics[scale=0.15]{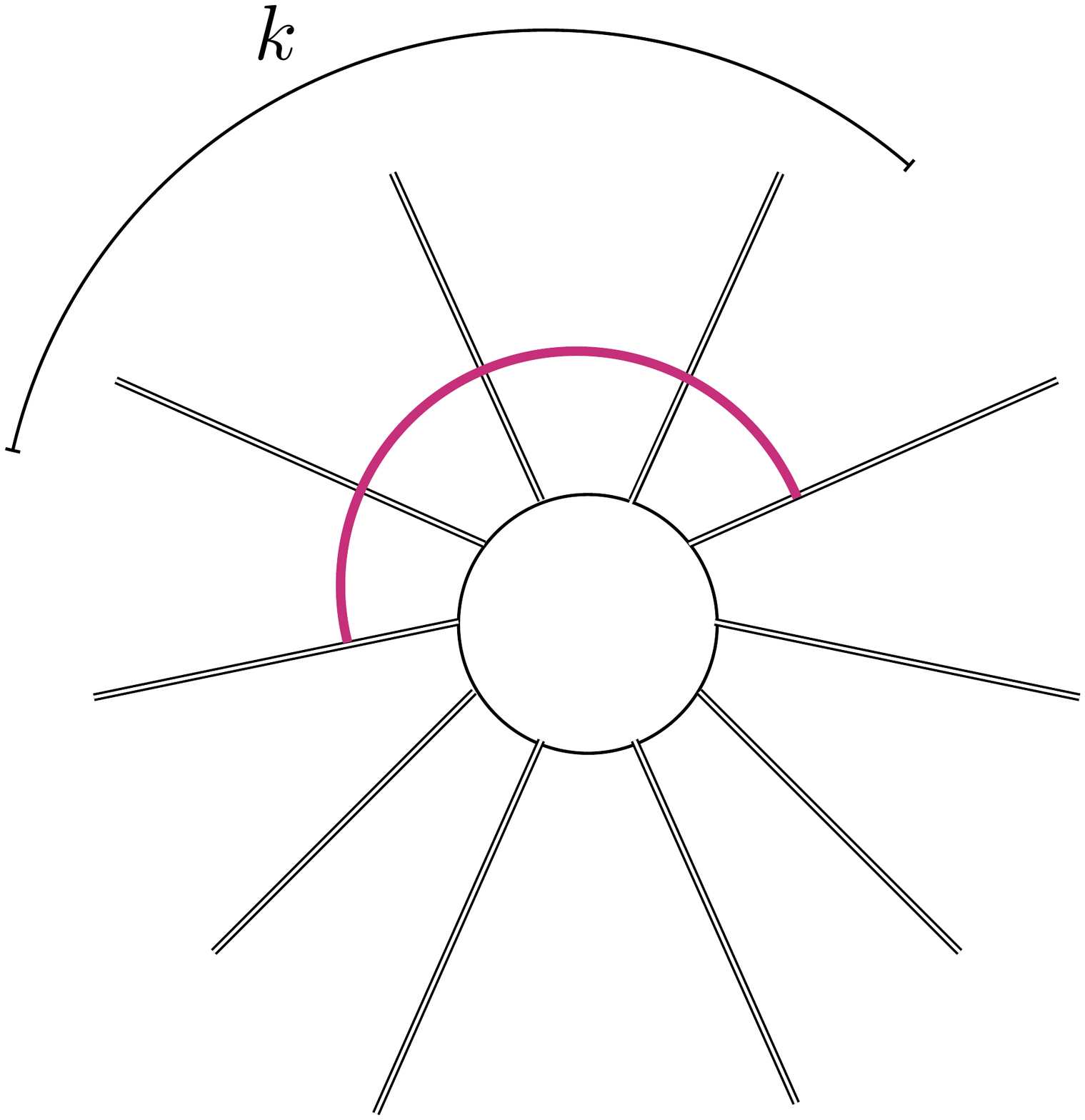}}} \label{actionM} 
\end{align}
 and obtain a recursion rule constraining the $g_{n,k}$'s to ensure absence of tree level particle production in the planar limit. As in the previous subsection, we will only discuss necessary conditions coming from cancellation of particle production in the multi-Regge limit, and omit an analysis of the seed for the recursion, i.e. analogues of \ref{sec:SGSingularityStruct} and \ref{sec:BaseBD}.
 
First, we will systematically fix field redefinition ambiguities in (\ref{actionM}). We shall then discuss the expected scattering behavior in integrable massless theories. Finally, we define the appropriate multi-regge limit and construct the recursion relation. It admits a single solution - the {$U(N)$} Non-Linear-Sigma-Model (NLSM).

\subsubsection{Amplitudes, kinematics and jets} 
\label{sec:ampsAndSuch}
We decompose the full amplitude according to the trace structure and focus on the single-trace parts 
\begin{equation}
	\mathcal{M}^{\tau_1\dots \tau_k}\left(p_1,\dots,p_k\right) \supset \sum_{\sigma \in S_{n}/Z_n} Tr\left[T^R_{\tau_{\sigma(1)}}\dots T^R_{\tau_{\sigma(n)}}\right]\mathcal{M}\left(p_{\sigma(1)},\dots,p_{\sigma(k)}\right)\quad \,,
\end{equation}
where $\mathcal{M}$ is a planar ordered amplitude. Vanishing of the full amplitude implies the vanishing of the planar ordered amplitude. In two dimensions, when we scatter massless particles they can be right or left movers with \begin{equation}
p_{+} \equiv \{p,0\} \quad,\quad\text{and}\quad,\quad p_-\equiv \{0,p\} \quad,
\end{equation}
respectively so these partial amplitudes will split further into a bunch of independent possibilities as 
\begin{equation}
\mathcal{M}(-++-)= \mathcal{M}(p_-^{(1)},p_+^{(2)},p_+^{(3)},p_-^{(4)}) \,, \qquad \mathcal{M}(-+-+)= \mathcal{M}(p_-^{(1)},p_+^{(2)},p_-^{(3)},p_+^{(4)})\,, \,\,\,  \text{etc}\,. \label{examples}
\end{equation}
Note that because of the left/right moving nature of massless particles in two dimensions, multi-particle scattering involving large number of particles can have a dramatically different space-time interpretation. Take an extreme example such as 
\begin{equation}
\mathcal{M}(\underbrace{----+++++}_\text{incoming}\underbrace{++++-----}_\text{outgoing}) \label{dangerous}
\end{equation}
which would describe a collision of two big jets of collinear particles to produce two jets in the final state. Since jets resemble individual particles one could expect that this big amplitude would be simply proportional to the first $2\to 2$ amplitude in (\ref{examples}) with some large momenta and be non-zero even in an integrable theory. In the other extreme, an amplitude like 
\begin{equation}
\mathcal{M}(\underbrace{-+-+-+-+}_\text{incoming}\underbrace{-+-+-+-+}_\text{outgoing}) \label{nice}
\end{equation}
would describe a bunch of non-collinear particles colliding into another non-degenerate bunch. In an integrable theory we expect this to vanish. Another example would be 
\begin{equation}
\mathcal{M}(\underbrace{----+++++}_\text{incoming}\underbrace{++++-----+++++}_\text{outgoing})
\end{equation}
which would describe two jets colliding into three jets and which we would again expect to vanish in an integrable theory. (and be proportional to the five particle amplitude in a non-integrable theory)  To summarize, when cancelling particle production we want to impose that all amplitudes involving many particles are zero except, potentially, the dangerous case~(\ref{dangerous}). Nicely, we will see below that imposing the cancellation of the most non-degenerate scattering configurations such as (\ref{nice}) (and of small deformations thereof) is already enough to completely constrain all the couplings and allow us to rediscover the NLSM as the unique massless matrix valued theory with two derivative interactions and no tree-level planar particle production. 

\subsubsection{Field redefinitions}

Under the field redefinition $h\to h+\alpha_3 h^2+\alpha_4 h^3+\dots$ we obtain a new Lagrangian of the same form as in (\ref{actionM}) but with the couplings $g_{n,k}$ reshuffled. More precisely, $\alpha_3$ shifts the cubic couplings ($n=3$) \textit{and higher}, $\alpha_4$ affects the  quartic couplings $(n=4)$ and higher etc in this triangular fashion. So we can exploit this field redefinition freedom to set to zero one of the couplings $g_{n,k}$ at each $n$ for example. We use it to set 
\begin{equation}
g_{n,2}=0 \,. \label{gauge}
\end{equation} 

At this point, this choice could seem rather arbitrary as we can also make other choices such as taking $g_{n,1}$ to be zero or any other more complicated choices of which $k$'s to contraints for each $n$ but we will see below that the $g_{n,2}=0$ has great advantage when we apply the induction process. 

Note also that at the cubic level $g_{3,1}=g_{3,2}$ so we are killing the cubic coupling altogether and we see that can restrict to theories without cubic couplings without any loss of generality. 

Since we have no cubic coupling and since we set one of the two independent quartic couplings to zero, we see that the four-particle scattering at tree level is given by a single quartic interaction $\mathcal{L}_\text{quartic}=2 g_{4,1} \text{Tr}(h^2 (\partial h)^2)$ so that $\mathcal{M}(p_1,\dots,p_4) \propto (p_1\cdot p_2)+ (p_2\cdot p_3) \propto (p_2 \cdot p_4)$ using momentum conservation and massless conditions. Hence, the second amplitude in the example list (\ref{examples}) vanishes while the first one, $\mathcal{M}(-++-)$ survives.
%
%
%

\subsubsection{Vanishing of odd terms}

\begin{figure}[t]
\centering
\includegraphics[scale=0.4]{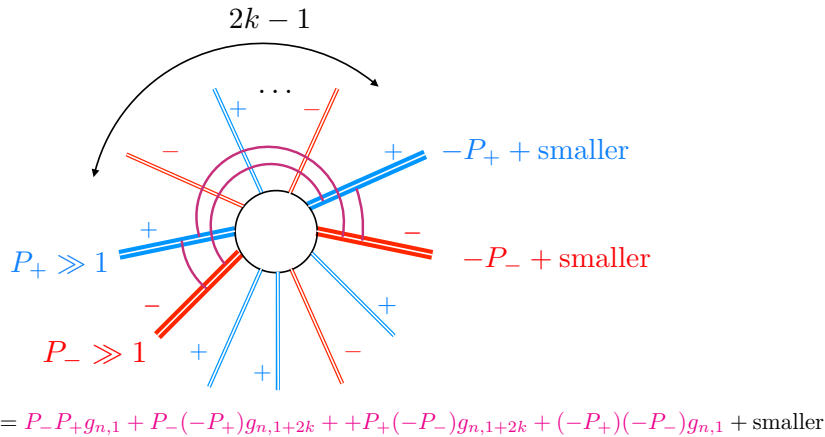}
\vspace{-2.4cm}
\caption{If four momenta are much larger than the other ones then the contact amplitude is simply given by the four terms in the vertex which couple pairs of such large momenta with opposite chirality. No particle production thus sets those simple combinations to vanish. Carefully choosing which momenta are large immediately lead to $g_{n,2k+1}=g_{n,1}$ as illustrated here. 
\label{illustrationOdd}
}
\end{figure}

Since we have no cubic couplings, the five particle scattering process is purely given by contact interactions given by the five particle vertices. Imposing that these vanish set all the quintic couplings to zero.Then the seven particle scattering process is again a purely contact interaction and setting it to vanish again sets all $n=7$ couplings to vanish and so on. To see this rather explicitly, consider for example the amplitude 
\begin{equation}
\mathcal{M}(--+-+-+-\dots)
\end{equation}
and take two of the left-moving momenta (and two of the right-moving momenta) to be very large, much larger than all other momenta, and with opposite sign as to be compatible with momentum conservation. Then, to leading order in the magnitude of the momenta of these very energetic particles, the amplitude -- which is given by contact interactions only -- is given by the only vertices whose derivatives couple these four very large momenta. As illustrated in figure \ref{illustrationOdd}, by playing with which momenta we take to be large we can in this way readily show that 
\begin{equation}
g_{n,2k+1}=g_{n,1}=g_{n}^\text{odd} \label{gOdd}
\end{equation}
and since the cyclic relation $g_{n,k}=g_{n,n-k}$ relates $k$ and $n-k$ which have different parity, we conclude that $g_{n,k}$ is actually $k$ independent altogether. Since we have already $g_{n,2}=0$ from our gauge choice (\ref{gauge}) we thus conclude that 
\begin{equation}
g_{n,k}=0 \qquad \text{for } n \text{ odd} \,,
\end{equation}
for theories without particle production.

\subsubsection{Determination of even terms}
Next we move to the even couplings. We first impose the vanishing of the alternating and nearly alternating amplitudes 
\begin{equation}
\mathcal{M}(+-+\dots-+{\color{red}-}) \,, \qquad \mathcal{M}(+-+\dots-+{\color{red}+}) \label{alternatingM} \,.
\end{equation}
To do so, we will use the vanishing of the more general off-shell \textit{currents} 
\begin{equation}
\mathcal{M}(+-+\dots-+{\color{red}\alpha}) \label{currents}
\end{equation} 
where the last particle is off-shell. It is clear that the first of these currents vanishes, $\mathcal{M}(+-+\alpha) = -4\,g_{4,1}\,p_1\cdot p_3 = 0$. If such currents vanish for $n$ particles or less then the same currents for $n+2$ particles are given by contact vertices only since any internal propagator will have currents with $n$ particles or less on one of it's sides. Hence, our full induction loop goes as follows

\begin{figure}[t]
	\centering
	\includegraphics[scale=0.45]{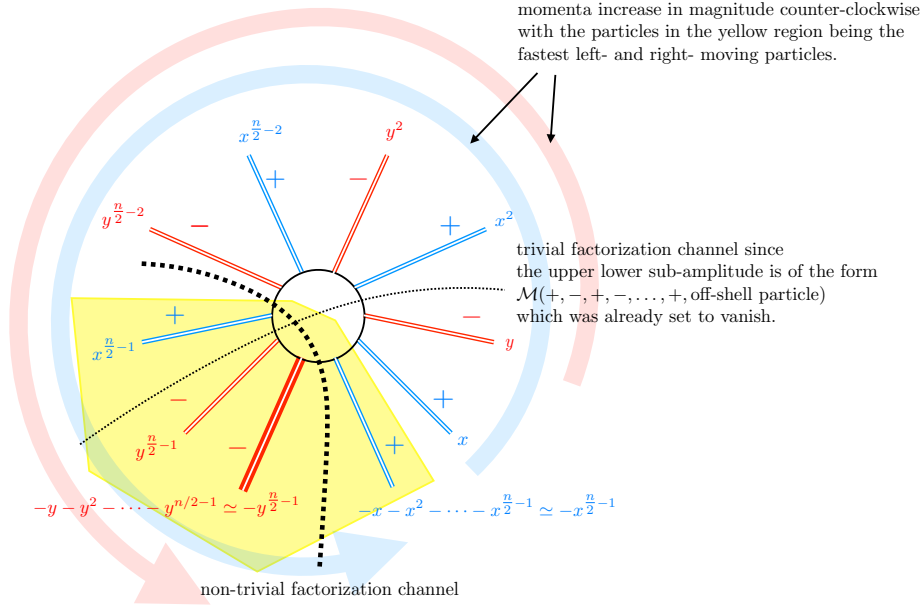}
	\vspace{-0.8cm}
	\caption{Non-vanishing factorizations have $p_1$ and $p_n$ on different sides of the propagator. Organizing the contributions by the vertex including $p_1$ (see figure \ref{fig:even-channels}) we see that all of them are proportional to full amplitudes. Hence, only the one with an $(n-2)$-particle vertex on one side and the $4$-particle amplitude on the other survives. \label{even-illustration}
	}
\end{figure}

\begin{enumerate}
\item Start with amplitudes (\ref{alternatingM}) \textit{and} currents which vanish for $n$ particles.
\item The amplitudes of the form $n+2$ are given by pure contact vertices since in their factorization channels only vanishing lower amplitudes such as (\ref{alternatingM}) and currents of the form (\ref{currents}) show up. Imposing that these contact amplitudes vanish mimics the odd $n$ analysis of the previous section almost verbatim. A key difference here is that $g_{n,k}=g_{n,n-k}$ now relates $k$ and $n-k$ which have the same parity so the even and odd terms are more independent now. Indeed, following the very same limits as in figure \ref{illustrationOdd} for the amplitudes (\ref{alternatingM}) immediately leads to 
\begin{equation}
g_{n,k} = \delta_{n \text{ even}} \delta_{k \text{ odd} } \, g_{n} \,.
\end{equation}
\item Finally we check that with these couplings the more general currents (\ref{currents}) also vanish for $n+2$ particles. This is a rather straightforward exercise since for purely contact interactions there is no big difference between currents and amplitudes. Explicit computation indeed leads to a vanishing result for these currents. We thus have a perfect induction loop. 
\end{enumerate}
We will now derive a simple recursion relation on these $g_{n}$. We impose the vanishing of the $n$-particle amplitude,
\begin{equation}
	\mathcal{M}(++-\dots+--)
\end{equation}
in the limit,
\begin{equation}\label{eq:NLSMlimit2}
p_{j\text{-even}} = x^{j/2}\qquad,\qquad p_{j\text{-odd}} = y^{(j-1)/2}\qquad.
\end{equation}
$p_{1}$ and $p_{n}$ are determined by momenta conservation. Channels that have $p_1$ and $p_{n}$ on the same side of the propagator must vanish (see figure \ref{even-illustration}) since there is a  vanishing current (\ref{currents}) on the other side of the propagator.

\begin{figure}[t]
	\centering
  \includegraphics[scale=0.45]{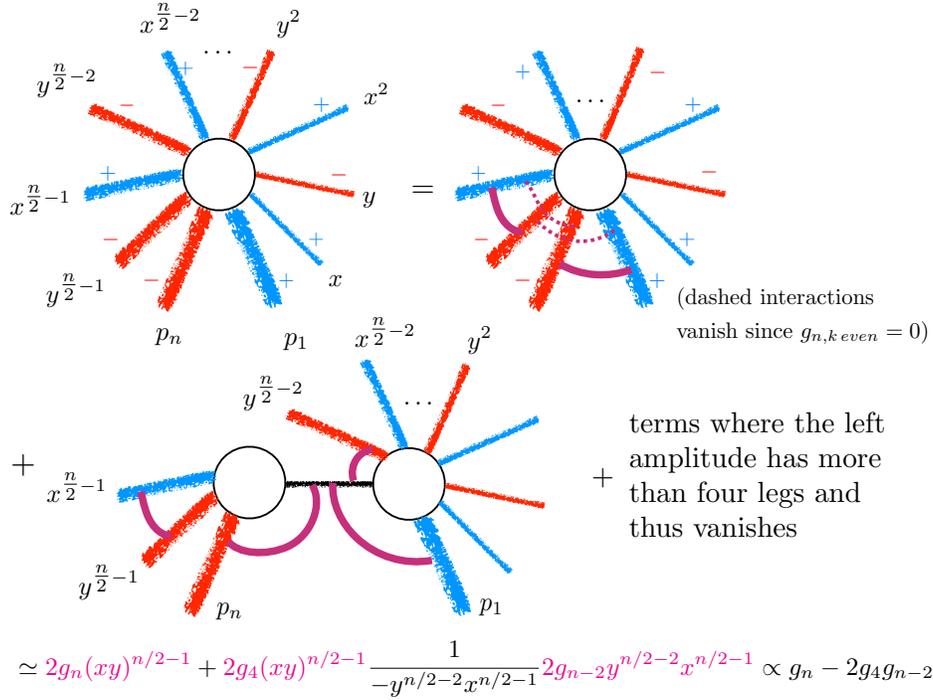}
	\caption{As done for the case with no derivative interactions, one groups the Feynman diagrams contributing to this amplitude by the vertex containing $p_1$. The sum of diagrams in the group corresponding to an $m$-particle vertex is proportional, to leading order, to a full $(n-m+2)$-particle amplitude and vanishes for $m<n-2$. The propagator can be treated as an \textit{on-shell} right moving particle because its left-moving part is of order $y^{m/2-1}$ and is vanishingly small in comparison to any left-moving momenta to the left of the propagator. \label{fig:even-channels}
	}
\end{figure}

The resulting constraint is depicted in figure \ref{fig:even-channels} leading to
\begin{equation}
	g_{n} = 2g_{4}\,g_{n-2} \quad. \label{recursionUN}
\end{equation}
leading to $g_n=\frac{1}{2} F^{-n+2}$ with $F$ being a constant, i.e. 
\begin{equation}
	g_{n,k} = \frac{1+(-1)^n}{2}\frac{1-(-1)^k}{2} \frac{1}{2} F^{-n+2}\, \,. \label{finalSol}
	\end{equation}
In the next section we identify a well known theory corresponding precisely to these couplings. 

\subsubsection{The Cayley parametrization of the non-linear sigma model} 
\label{sec:NLSMCayley}

In the last section we found the only candidate for integrability defined by a Lagrangian of the form (\ref{actionM}), in other words we proved it's uniqueness. However, we did not prove that all amplitudes of more than four jets (see section \ref{sec:ampsAndSuch}) vanish. In other words, we did not prove existence. Since it is well known that the $U(N)$ non-Linear-Sigma-Model (NLSM) is a quantum integrable theory, it better be that the unique solution we found corresponds precisely to this well known theory! This is what we verify in this section.  

To do so we will simply re-sum our Lagrangian. Plugging (\ref{finalSol}) into (\ref{actionM}) we get 
\begin{align}
\mathcal L(g)=\frac 1 2\sum _{a,b\ge 0}\frac{1+(-1)^a}{2}\frac{1+(-1)^b}{2}\frac{1}{F^{a+b}}\text{Tr}\left[ {h^{a}(\partial _{\mu}h)h^{b}(\partial ^{\mu}h)}\right]
\end{align}
which we can re-sum into 
\begin{align}
\mathcal L(g)=\frac 1 2 \text{Tr}\left[ \frac{1}{1-h^2/F^2}{(\partial _{\mu}h)  \frac{1}{1-h^2/F^2} (\partial ^{\mu}h)}\right]
\end{align}
At this point we note that the objects showing up are the derivatives of an $U(N)$ group element 
\begin{align}
\partial_\mu\left(g \equiv \frac{1+h/F}{1-h/F} \right) =\partial_\mu \left(1+2\sum _{n=1}^{\infty}\frac {h^n}{F^{n}} \right)= \frac{1}{2}\sum_{a,b \ge 0} \frac{h^a (\partial_\mu h) h^b}{F^{a+b+1}} =\frac{1}{2F}\frac{1}{1-h/F} \partial_\mu h \frac{1}{1-h/F} \,. \label{parG}
\end{align}
Indeed, $g^{-1}$ is equal to the group element $g$ with $h\to -h$ and hence, using cyclicity of the trace, we see that 
\begin{align}
\mathcal L(g)=\frac{1}{2} \text{Tr}\Big[\underbrace{ \frac{1}{1-h/F}{(\partial _{\mu}h)  \frac{1}{1-h/F}}}_{2F\partial_\mu g} \underbrace{\frac{1}{1+h/F} (\partial^{\mu}h) \frac{1}{1+h/F} }_{2F\partial_\mu g^{-1}}\Big] = 2F^2 \big[\text{Tr}(\partial_\mu g)(\partial^\mu g^{-1}) \big] 
\end{align}
thus precisely recognizing the NLSM as expected! The parametrization of $g$ in (\ref{parG}) is known as the Cayley parametrization, see e.g. a recent work \cite{Kampf:2013vha} exploring these and many other parametrizations in a higher-dimensional context. 
%
%

\section{Summary}
In this paper, we considered relativistic two-dimensional quantum field theories at tree level. We investigated how imposing the absence of particle production can efficiently restrict the Lagrangians of these theories, often determining them completely. Key in our analysis was the idea of using high-energy limits to isolate particular exchange processes and thus tame the otherwise very complicated tree-level combinatorics. The chief example is the multi-Regge limit introduced in Figure \ref{LimitFig}. Using these high-energy limits, we derived recursion relations constraining various couplings in theories with no particle production. We found (\ref{recursion}) for theories with a single massive scalar, (\ref{eq:recursionN}) for multiple massive scalars and (\ref{recursionUN}) for theories with massless matrix-valued fields with two-derivative interactions. By solving these recursion relations, we made contact with well-known integrable theories such as sine-Gordon, Bullough-Dodd, multiple Toda theories and the non-linear sigma models.\footnote{For the later cases, we derived the recursion relations but did not fully analyse the seed and induction step which we did in section \ref{sec:miracle} for the single scalar case. It would be important and illuminating to close this gap if we are to look for new models in this way. It would also be interesting to think of using such limits at the quantum level.} (Especially for the multiple field theories) 
it would be very interesting to perform a more systematic analysis of the recursion relations and seeds. It would be formidable if we could unveil new overlooked integrable models as solutions to these high-energy recursion relations.

\section*{Acknowledgments}
We thank Benjamin Basso, Freddy Cachazo, Lucia Cordova, Patrick Dorey, Alexandre Holmrich, Joao Penedones, Jon Toledo and Alexander Zamolodchikov for numerous enlightening discussions and suggestions. {BG is especially grateful to Freddy Cachazo for devoted supervision of his PSI thesis, which followed lines similar to this work}. Research at the Perimeter Institute is supported in part by the Government of Canada through NSERC and by the Province of Ontario through MRI. 
This research received funding from the grant CERN/FIS-NUC/0045/2015.
This work was additionally supported by a grant from the Simons Foundation (PV: \#488661). PV thanks FAPESP grant 2016/01343-7 for partial financial support.

\appendix

\begingroup
\let\cleardoublepage\clearpage
\bibliographystyle{utphys}
\bibliography{references} 
\endgroup
\clearpage


\end{document}

%% file: sumToRecur.pdf_tex
\begingroup%
  \makeatletter%
  \providecommand\color[2][]{%
    \errmessage{(Inkscape) Color is used for the text in Inkscape, but the package 'color.sty' is not loaded}%
    \renewcommand\color[2][]{}%
  }%
  \providecommand\transparent[1]{%
    \errmessage{(Inkscape) Transparency is used (non-zero) for the text in Inkscape, but the package 'transparent.sty' is not loaded}%
    \renewcommand\transparent[1]{}%
  }%
  \providecommand\rotatebox[2]{#2}%
  \ifx\svgwidth\undefined%
    \setlength{\unitlength}{1644.02372586bp}%
    \ifx\svgscale\undefined%
      \relax%
    \else%
      \setlength{\unitlength}{\unitlength * \real{\svgscale}}%
    \fi%
  \else%
    \setlength{\unitlength}{\svgwidth}%
  \fi%
  \global\let\svgwidth\undefined%
  \global\let\svgscale\undefined%
  \makeatother%
  \begin{picture}(1,0.35122356)%
    \put(0,0){\includegraphics[width=\unitlength,page=1]{sumToRecur.pdf}}%
    \put(0.03937323,0.26340943){\color[rgb]{0,0,0}\makebox(0,0)[lb]{\smash{$2$}}}%
    \put(0.08532215,0.21578085){\color[rgb]{0,0,0}\makebox(0,0)[lb]{\smash{$1$}}}%
    \put(0,0){\includegraphics[width=\unitlength,page=2]{sumToRecur.pdf}}%
    \put(0.37093072,0.26204599){\color[rgb]{0,0,0}\makebox(0,0)[lb]{\smash{$2$}}}%
    \put(0.41687963,0.2144174){\color[rgb]{0,0,0}\makebox(0,0)[lb]{\smash{$1$}}}%
    \put(0.30777646,0.26077138){\color[rgb]{0,0,0}\makebox(0,0)[lb]{\smash{$+$}}}%
    \put(0.03258963,0.07715925){\color[rgb]{0,0,0}\makebox(0,0)[lb]{\smash{$+$}}}%
    \put(0,0){\includegraphics[width=\unitlength,page=3]{sumToRecur.pdf}}%
    \put(0.08126059,0.07684804){\color[rgb]{0,0,0}\makebox(0,0)[lb]{\smash{$2$}}}%
    \put(0.12526308,0.03019268){\color[rgb]{0,0,0}\makebox(0,0)[lb]{\smash{$1$}}}%
    \put(0.09217354,0.10999687){\color[rgb]{0,0,0}\makebox(0,0)[lb]{\smash{$3$}}}%
    \put(0.2605409,0.14405963){\color[rgb]{0,0,0}\makebox(0,0)[lb]{\smash{$n$}}}%
    \put(0.30628233,0.07788055){\color[rgb]{0,0,0}\makebox(0,0)[lb]{\smash{$(n+1)$}}}%
    \put(0.24496935,0.0117015){\color[rgb]{0,0,0}\makebox(0,0)[lb]{\smash{$(n+2)$}}}%
    \put(0.40476057,0.0780436){\color[rgb]{0,0,0}\makebox(0,0)[lb]{\smash{$+$}}}%
    \put(0,0){\includegraphics[width=\unitlength,page=4]{sumToRecur.pdf}}%
    \put(0.64310105,0.14562576){\color[rgb]{0,0,0}\makebox(0,0)[lb]{\smash{$(n+1)$}}}%
    \put(0.64245388,0.01267481){\color[rgb]{0,0,0}\makebox(0,0)[lb]{\smash{$(n+2)$}}}%
    \put(0.60822396,0.26174459){\color[rgb]{0,0,0}\makebox(0,0)[lb]{\smash{$+\ \,\dots\ \,+$}}}%
    \put(0,0){\includegraphics[width=\unitlength,page=5]{sumToRecur.pdf}}%
    \put(0.29156615,0.31891273){\color[rgb]{0,0,0}\makebox(0,0)[lb]{\smash{$0$}}}%
    \put(0,0){\includegraphics[width=\unitlength,page=6]{sumToRecur.pdf}}%
    \put(0.62411651,0.3190955){\color[rgb]{0,0,0}\makebox(0,0)[lb]{\smash{$0$}}}%
    \put(0.41590641,0.30687347){\color[rgb]{0,0,0}\makebox(0,0)[lb]{\smash{$3$}}}%
    \put(-0.04800925,-0.02265852){\color[rgb]{0,0,0}\makebox(0,0)[lt]{\begin{minipage}{0.79804191\unitlength}\raggedright \end{minipage}}}%
    \put(0.70599011,0.08009872){\color[rgb]{0,0,0}\makebox(0,0)[lb]{\smash{$+$}}}%
    \put(0.11716926,0.10011074){\color[rgb]{0,0,0}\makebox(0,0)[lb]{\smash{$\dots$}}}%
    \put(0,0){\includegraphics[width=\unitlength,page=7]{sumToRecur.pdf}}%
    \put(0.45859262,0.07786926){\color[rgb]{0,0,0}\makebox(0,0)[lb]{\smash{$2$}}}%
    \put(0.51427378,0.03218712){\color[rgb]{0,0,0}\makebox(0,0)[lb]{\smash{$1$}}}%
    \put(0.48118421,0.11101809){\color[rgb]{0,0,0}\makebox(0,0)[lb]{\smash{$3$}}}%
    \put(0.54349021,0.11445893){\color[rgb]{0,0,0}\makebox(0,0)[lb]{\smash{$n$}}}%
    \put(0.50617998,0.10113196){\color[rgb]{0,0,0}\makebox(0,0)[lb]{\smash{$\dots$}}}%
    \put(0,0){\includegraphics[width=\unitlength,page=8]{sumToRecur.pdf}}%
    \put(0.75556431,0.07796029){\color[rgb]{0,0,0}\makebox(0,0)[lb]{\smash{$2$}}}%
    \put(0.81319188,0.03130493){\color[rgb]{0,0,0}\makebox(0,0)[lb]{\smash{$1$}}}%
    \put(0.78302201,0.11597523){\color[rgb]{0,0,0}\makebox(0,0)[lb]{\smash{$3$}}}%
    \put(0.83364932,0.11552319){\color[rgb]{0,0,0}\makebox(0,0)[lb]{\smash{$(n+1)$}}}%
    \put(0.80509806,0.10122299){\color[rgb]{0,0,0}\makebox(0,0)[lb]{\smash{$\dots$}}}%
    \put(0,0){\includegraphics[width=\unitlength,page=9]{sumToRecur.pdf}}%
    \put(0.87235093,0.07893352){\color[rgb]{0,0,0}\makebox(0,0)[lb]{\smash{$(n+2)$}}}%
    \put(0,0){\includegraphics[width=\unitlength,page=10]{sumToRecur.pdf}}%
    \put(0.09217354,0.10999687){\color[rgb]{0,0,0}\makebox(0,0)[lb]{\smash{$3$}}}%
    \put(0.1505866,0.11538416){\color[rgb]{0,0,0}\makebox(0,0)[lb]{\smash{$(n-1)$}}}%
    \put(0.11716926,0.10011074){\color[rgb]{0,0,0}\makebox(0,0)[lb]{\smash{$\dots$}}}%
    \put(0,0){\includegraphics[width=\unitlength,page=11]{sumToRecur.pdf}}%
    \put(0.5744506,0.28416354){\color[rgb]{0,0,0}\rotatebox{-90}{\makebox(0,0)[lb]{\smash{$\dots$}}}}%
    \put(0.24444936,0.28502808){\color[rgb]{0,0,0}\rotatebox{-90}{\makebox(0,0)[lb]{\smash{$\dots$}}}}%
    \put(0,0){\includegraphics[width=\unitlength,page=12]{sumToRecur.pdf}}%
    \put(0.73699967,0.26124768){\color[rgb]{0,0,0}\makebox(0,0)[lb]{\smash{$2$}}}%
    \put(0.78100214,0.21459232){\color[rgb]{0,0,0}\makebox(0,0)[lb]{\smash{$1$}}}%
    \put(0.7479126,0.2943965){\color[rgb]{0,0,0}\makebox(0,0)[lb]{\smash{$3$}}}%
    \put(0.77290832,0.28451038){\color[rgb]{0,0,0}\makebox(0,0)[lb]{\smash{$\dots$}}}%
    \put(0,0){\includegraphics[width=\unitlength,page=13]{sumToRecur.pdf}}%
    \put(0.7479126,0.2943965){\color[rgb]{0,0,0}\makebox(0,0)[lb]{\smash{$3$}}}%
    \put(0.80048636,0.29783735){\color[rgb]{0,0,0}\makebox(0,0)[lb]{\smash{$(n-2)$}}}%
    \put(0.77290832,0.28451038){\color[rgb]{0,0,0}\makebox(0,0)[lb]{\smash{$\dots$}}}%
    \put(0,0){\includegraphics[width=\unitlength,page=14]{sumToRecur.pdf}}%
    \put(0.98526082,0.32451622){\color[rgb]{0,0,0}\makebox(0,0)[lb]{\smash{$0$}}}%
    \put(0,0){\includegraphics[width=\unitlength,page=15]{sumToRecur.pdf}}%
    \put(0.95073045,0.28502823){\color[rgb]{0,0,0}\rotatebox{-90}{\makebox(0,0)[lb]{\smash{$\dots$}}}}%
  \end{picture}%
\endgroup%

%% file: NoParticleProduction_v6.bbl
\providecommand{\href}[2]{#2}\begingroup\raggedright\begin{thebibliography}{1}

\bibitem{bootstrap1}
M.~F. Paulos, J.~Penedones, J.~Toledo, B.~C. van Rees, and P.~Vieira, ``{The
  S-matrix bootstrap II: two dimensional amplitudes},''
  \href{http://dx.doi.org/10.1007/JHEP11(2017)143}{{\em JHEP} {\bfseries 11}
  (2017) 143},
\href{http://arxiv.org/abs/1607.06110}{{\ttfamily arXiv:1607.06110 [hep-th]}}.

\bibitem{bootstrap2}
M.~F. Paulos, J.~Penedones, J.~Toledo, B.~C. van Rees, and P.~Vieira, ``{The
  S-matrix Bootstrap III: Higher Dimensional Amplitudes},''
\href{http://arxiv.org/abs/1708.06765}{{\ttfamily arXiv:1708.06765 [hep-th]}}.

\bibitem{Dorey:1996gd}
P.~Dorey, ``{Exact S matrices},''
\href{http://arxiv.org/abs/hep-th/9810026}{{\ttfamily arXiv:hep-th/9810026
  [hep-th]}}.

\bibitem{Arefeva:1974bk}
I.~Arefeva and V.~Korepin, ``{Scattering in two-dimensional model with
  Lagrangian (1/gamma) ((d(mu)u)**2/2 + m**2 cos(u-1))},''
{\em Pisma Zh. Eksp. Teor. Fiz.} {\bfseries 20} (1974) 680.

\bibitem{Braden:1991vz}
H.~W. Braden and R.~Sasaki, ``{Affine Toda perturbation theory},''
\href{http://dx.doi.org/10.1016/0550-3213(92)90601-7}{{\em Nucl. Phys.}
  {\bfseries B379} (1992) 377--428}.

\bibitem{brotherPaper}
C.~Bercini and D.~Trancanelli, ``{Supersymmetric integrable models from no
  particle production},'' {\em To Appear} (2018) .

\bibitem{ATFT-Mussardo}
G.~Mussardo, ``{Off critical statistical models: Factorized scattering theories
  and bootstrap program},''
\href{http://dx.doi.org/10.1016/0370-1573(92)90047-4}{{\em Phys. Rept.}
  {\bfseries 218} (1992) 215--379}.

\bibitem{Kampf:2013vha}
K.~Kampf, J.~Novotny, and J.~Trnka, ``{Tree-level Amplitudes in the Nonlinear
  Sigma Model},'' \href{http://dx.doi.org/10.1007/JHEP05(2013)032}{{\em JHEP}
  {\bfseries 05} (2013) 032},
\href{http://arxiv.org/abs/1304.3048}{{\ttfamily arXiv:1304.3048 [hep-th]}}.

\end{thebibliography}\endgroup
